%% file: p-modeSouth.tex
\documentclass{aa}

\usepackage[T1]{fontenc}
\usepackage{ae,aecompl}
\usepackage{xtab}
\usepackage{caption}
\usepackage{courier}
\usepackage{multirow}
\usepackage{times}
\usepackage{upgreek}
\usepackage{graphicx}
\usepackage{epstopdf}
\usepackage{newtxtext}
\usepackage{newtxmath}
\usepackage{amsmath}
\usepackage{amssymb}
\usepackage[section]{placeins}
\usepackage{tablefootnote}
\usepackage{color}
\usepackage{gensymb}
\usepackage{longtable}
\usepackage{natbib}

\newcommand{\uHz}{$\upmu$Hz}
\newcommand{\teff}{$T_{\rm eff}$}
\newcommand{\logg}{$\log$\,g}
\newcommand{\loggcms}{$\log{(g/\cmss)}$}
\newcommand{\kep}{{\it Kepler}}
\newcommand{\ktwo}{{\it K2}}

\newcommand{\tess}{{\it TESS}}
\newcommand{\corot}{{\it CoRoT}}

\newcommand{\kmsec}{\,\mbox{$\mbox{km}\,\mbox{s}^{-1}$}}    
\newcommand{\cmss}{\mbox{$\mbox{cm}\,\mbox{s}^{-2}$}}    

\begin{document} 

\title{Short-period pulsating hot-subdwarf stars observed by TESS}
\subtitle{I. Southern ecliptic hemisphere\thanks{Table\,4 is only available in electronic form at the CDS via anonymous ftp to cdsarc.u-strasbg.fr (130.79.128.5) or via http://cdsweb.u-strasbg.fr/cgi-bin/qcat?J/A+A/}}
   \author{A.S.\,Baran\inst{1,2,3}, V.\,Van Grootel\inst{4}, R.H.\,{\O}stensen\inst{3,5}, H.L.\,Worters\inst{6}, S.K.\,Sahoo\inst{1,7}, S.\,Sanjayan\inst{1,7}, S.\,Charpinet\inst{8}, P.\,Nemeth\inst{9,10}, J.H.\,Telting\inst{11,12}, D.\,Kilkenny\inst{13}
   }

\institute{ARDASTELLA Research Group
    \and Astronomical Observatory, University of Warsaw, Al. Ujazdowskie 4, 00-478 Warszawa, Poland
    \and Department of Physics, Astronomy, and Materials Science, Missouri State University, Springfield, MO\,65897, USA
    \and Space sciences, Technologies and Astrophysics Research (STAR) Institute, Université de Liège, 19C Allée du 6 Août, B-4000 Liège, Belgium
    \and Recogito AS, Storgaten 72, N-8200 Fauske, Norway
    \and South African Astronomical Observatory, Observatory 7935, South Africa
    \and Nicolaus Copernicus Astronomical Centre of the Polish Academy of Sciences, ul. Bartycka 18, 00-716 Warsaw, Poland
    \and IRAP, CNRS, UPS, CNES, Université de Toulouse, 14 av. Edouard Belin, 31400 Toulouse, France
    \and Astronomical Institute of the Czech Academy of Sciences, Fri\v{c}ova 298, CZ-251\,65 Ond\v{r}ejov, Czech Republic
    \and Astroserver.org, F\H{o} t\'er 1, 8533 Malomsok, Hungary
    \and Nordic Optical Telescope, Rambla Jos{\'e} Ana Fern{\'a}ndez P{\'e}rez 7, 38711 Bre{\~n}a Baja, Spain
    \and Department of Physics and Astronomy, Aarhus University, NyMunkegade 120, DK-8000 Aarhus C, Denmark
    \and Department of Physics and Astronomy, University of the Western Cape, Private Bag X17, Bellville 7535, South Africa
}
\date{}


\abstract
{We present results of a Transiting Exoplanet Survey Satellite (\tess) search for short-period pulsations in compact stellar objects observed in years\,1 and 3 of the \tess\ mission, during which the southern ecliptic hemisphere was targeted. We describe the \tess\ data used and the details of the search method. For many of the targets, we use unpublished spectroscopic observations to classify the objects. From the \tess\ photometry, we clearly identify 43 short-period hot-subdwarf pulsators, including 32 sdB stars, eight sdOB stars, two sdO stars, and, significantly, one He-sdOB star, which is the first of this kind to show short-period pulsations. Eight stars show signals at both low and high frequencies, and are therefore ``hybrid'' pulsators. We report the list of prewhitened frequencies and we show the amplitude spectra calculated from the \tess\ data. We make an attempt to identify possible multiplets caused by stellar rotation, and we select four candidates with rotation periods between 1 and 12.9\,d. The most interesting targets discovered in this survey should be observed throughout the remainder of the \tess\ mission and from the ground. Asteroseismic investigations of these data sets will be invaluable in revealing the interior structure of these stars and will boost our understanding of their evolutionary history. We find three additional new variable stars but their spectral and variability types remain to be constrained.}


\authorrunning{A.S.\,Baran et al.}
\titlerunning{P-mode hot-subdwarf pulsators in the southern ecliptic hemisphere}
\maketitle

\section{Introduction}
\label{intro}
Hot subdwarf stars are evolved objects with O to B spectral types, and with a wide range of H to He fractions in their atmospheres. The subdwarf B (sdB) stars have effective temperatures, \teff, between 20\,000 and 40\,000\,K, surface gravities between 5 and 6 in \loggcms, and surface helium abundances usually strongly depleted compared to the solar value. The surface distribution of elements heavier than helium differs significantly from star to star \citep{geier13}. Most sdB stars are core-He burning stars that are located on the blue extension of the horizontal branch (EHB), and therefore have a mass of about half of the solar value, as required for the core-helium flash \citep{fontaine12}. The progenitors of sdB stars are intermediate-mass stars ($\sim$ 0.7-2.2\,M$_{\odot}$) that must have lost significant mass during the red-giant branch, leaving them with only a tiny remnant of their hydrogen envelopes. The mass loss must happen before helium ignition through the core-helium flash, otherwise they would become He-core low-mass white dwarfs (without He-core burning), or normal horizontal-branch stars (He-core burning, but with a much lower effective temperature, thicker H-rich envelopes, and higher masses). Binary population synthesis modeling has been performed to explore various mass-loss scenarios, as detailed by \cite{han02}. 

The subdwarf O (sdO) stars are hotter, showing \teff\ between $\sim$40\,000 and $\sim$80\,000\,K, and they exhibit a wider range of \loggcms,\ from 4.0 to 6.5. The majority have helium-enriched atmospheres. This diversity in properties betrays the very diverse origins of sdO stars. He-poor, compact sdO stars are direct progenies of sdB stars in their immediate post-EHB phase. However, the majority of compact sdO stars are He rich and are identified as direct post-RGB objects, that is to say they were created through the so-called late hot He-flash \citep{bertolami08}, or they are end products of merger events \citep{webbink84, iben84, iben90, saio00, saio02}. There are also less compact (\loggcms\ $\lesssim$ 5.0) sdO stars, which are post-AGB stars, that is to say stars that have ascended the asymptotic giant branch after core-He burning exhaustion \citep{reindl16}.

Stellar oscillations in sdB stars were predicted by \citet{charpinet96,charpinet97}, and  and discovered and independently discovered \cite{kilkenny97a}, and are potentially useful for precise sdB mass estimations using asteroseismology, among other applications. Pulsating sdB stars reported by \cite{kilkenny97a}, and in subsequent papers of their series, are pressure- (p-) mode pulsators. Later, gravity- (g-) mode sdB stars and sdB stars exhibiting both p and g modes (hybrid)  were reported by \citet{green03}, \citet{baran05}, and \citet{schuh06}. \cite{fontaine12} compared p- and g-mode sdB stars, clearly indicating that the typical periods of p-mode sdB stars are shorter, while amplitudes of the flux variation are higher than those of g-mode stars. Although the first pulsating subdwarf stars discovered were p-mode dominated, most of them stars turned out to be g-mode dominated. \citet{ostensen10} found that only 10\% of temperature-selected V361-Hya candidates in the Nordic Optical Telescope (NOT) survey are pulsators, while \citet{green03} found that 75\% of the cooler sdB stars, having \teff\,<30\,000\,K (or 25\%–30\% of all sdB stars), pulsate in g modes. The majority of sdB stars detected using the Convection, Rotation and planetary Transits (\corot), the \kep\, and the Transiting Exoplanet Survey Satellite (\tess)\ space telescopes are either g-mode pulsators or hybrid pulsators with a dominant g-mode component.

The first pulsating sdO star, V499\,Ser (=\,J1600+0748), was identified by \citet{woudt06}. It shows very rapid oscillations from roughly 60\,s to 120\,s. For years V499\,Ser remained the only member of its class, despite extensive searches among field sdO stars \citep{rodriguezlopez07,johnson14}. Four sdO pulsators were identified by \citet{randall11} in the globular cluster $\omega$\,Cen and, over several years, a few more field sdO pulsators were identified, including EO\,Cet, formerly identified as a common-or-garden sdB pulsator \citep{ostensen12}. The periods of all known sdO pulsators are of a few minutes and are consistent with p-mode pulsations \citep{fontaine08}.

Another type of pulsating subdwarf star is the V366-Aqr, which are intermediate He-rich hot subdwarfs (iHe-sdOBs)\footnote{The He abundance is intermediate between He poor and He enhanced. He poor is defined as subsolar, i.e. $\log$(n(He)/$\log$\,(H))\,<\,-1, and He enhanced is higher than neutral, i.e. n(He)\,>\,n(H) or $\log$(n(He)/$\log$\,(H))\,>\,0. So the intermediate region is between neutral and solar, i.e. between 0 and -1 in $\log$(n(He)/n(H)).} that pulsate with periods on the order of an hour. V366\,Aqr (=\,LS\,IV$-$14$^\circ$116) was discovered by \citet{ahmad05}, and further examples of this class were only discovered recently; Feige\,46 by \citet{latour19} and PHL\,417 by \citet{ostensen20}. These pulsators belong to the group of heavy-metal subdwarfs that show unusual enrichment of elements such as zirconium and lead in their atmospheres.

The rarest hot-subdwarf pulsator is currently EC\,03089-6421 \citep{kilkenny19}, an sdO star with extremely rapid half-minute pulsations. This star might be a higher-gravity object in a post-He-burning stage, but a more detailed spectroscopic study is required to resolve this. 

The goal of this paper is to report our results of the p-mode-dominated hot-subdwarf stars surveyed by the \tess\ satellite. Since only the southern ecliptic hemisphere is completed using the ultra-short cadence required for rapid pulsations, our report is limited to that hemisphere. The northern ecliptic hemisphere survey, along with a statistical analysis on the presence of p-mode pulsators among hot subdwarfs, will be reported in a follow-up paper.

\section{\tess\ photometry}
\label{data}
\tess\ is deployed in an elliptical, 2:1 lunar synchronous orbit with a period of 13.7\,d. Each annual cycle of \tess\ observations is split into sectors lasting two orbits, or about 27\,d. The detector consists of four contiguous CCD cameras, each covering a 24$^\circ$\,x\,24$^\circ$ field of view (FoV), making up a 24$^\circ$\,x\,96$^\circ$ strip aligned along ecliptic latitude lines. During years\,1 and 2, the data were stored with the short cadence (SC), lasting 120\,s, and the long cadence (LC), lasting 1800\,s. Since year\,3 there has been an additional ultra-short cadence (USC) available, lasting 20\,s, while the LC was trimmed to 600\,sec. The number of targets observed in the USC and SC is limited, and varies from sector to sector. When observations of one sector have been completed, the instrument's FoV is shifted eastward by 27$^\circ$, naturally pivoting around the ecliptic pole. During years\,1-3, it took 13 sectors to pivot around one pole, then the FoV was shifted to the other hemisphere for the next cycle. As a result, the regions near the ecliptic poles are observed during every sector and are known as the continuous viewing zones of \tess. We downloaded all available data of our targets from the Barbara A. Mikulski Archive for Space Telescopes (MAST)\footnote{archive.stsci.edu}.

To probe the p-mode region in an amplitude spectrum, we are limited to either the SC or USC data. The Nyquist frequency of the former is 4\,166\,\uHz, while for the latter it is 25\,000\,\uHz. Our preference is to use the USC data, if available, with which the entire p-mode region is sampled. In the case of the SC data, the Nyquist frequency is in the middle of the p-mode region and aliasing becomes a serious issue. This issue can be resolved by the finite speed of light, which clearly separates the sub- and super-Nyquist regions in the amplitude spectrum, first discovered by \citet{baran12a}. The amplitudes and profiles of the peaks become different in these two regions, but the effect does not come into play for stars located in the \tess\ continuous viewing zones, namely those close to the ecliptic poles.

We used PDCSAP\_FLUX, which is corrected for on-board systematics and neighbors' contributions to the overall flux. We clipped fluxes at 4.5$\upsigma$ to remove outliers, de-trended long-term variations (on the order of days) with spline fitting , and calculated the amplitude of the flux variations using the relation A[ppt]=1000*({\rm flux}/{<{\rm flux}>}-1).

\section{Fourier analysis}
\label{fourier}
About 4000 hot subdwarfs (and hot-subdwarf candidates) have been observed by TESS in either the SC or USC mode. The list was assembled by members of Working Group 8 (WG8) of the TESS Asteroseismic Science Consortium (TASC)\footnote{https://tasoc.dk}.

Following a recent report by \cite{baran21}, we applied a detection threshold at 4.5 times the median noise level to both the SC and USC data sets, regardless of the data coverage. In the case of data from just one sector, such a threshold corresponds to a false alarm probability (FAP) of 5\% for the SC data and 14\% for the USC data. For FAP\,=\,0.1\%, the threshold would be around 5.5, and hence frequencies below this threshold should be considered tentative. Precise confidence levels for other numbers of sectors can be derived from Eq.\,5 in \citet{baran21}.

To search for a significant signal in amplitude spectra, we applied the FELIX tool \citep{charpinet10,zong16}. Firstly, an automatic search was done, looking for variations beyond 1\,500 \uHz\ in one sector, without combining multiple sectors (if a star was observed during more than one sector). Secondly, an individual check was carried out on the $\sim$100 candidates to see if detected variations were consistent with p-mode pulsations. We detected signals in the p-mode region in 43 targets. Twelve targets were recorded only in the SC mode, while 31 were recorded in the USC mode. We found ten new detections in SC data and 11 new detections in USC data, while the remaining two\,(SC) and 20\,(USC) hot-subdwarf pulsators were known prior to the \tess\ mission. We present the full list of p-mode hot subdwarfs in Table\,\ref{tab:targets_all}. The table also lists the spectral type (SpT) either from the literature or from recent spectroscopy. As is conventional for the classification of hot subdwarfs from low-resolution spectroscopy, we designated those stars with detectable \ion{He}{i}, or no \ion{He}{I} lines at all, with a B class. We designated those with detectable \ion{He}{ii} at 4686\,\AA\ with class OB, and those with strong \ion{He}{ii} and no \ion{He}{i} with class O \citep[for more details, see e.g.,][]{moehler90}. One star is He rich, and shows both \ion{He}{i} and \ion{He}{ii} at roughly equal strength, but not stronger than the Balmer lines, indicating a super-solar helium abundance, which would imply a designation of He-sdOB or iHe-sdOB, as the atmosphere is still Hydrogen dominated. Below we describe each target, providing prior knowledge on the pulsation properties, followed by amplitude spectra and the lists of frequencies we detected.

\input{targets_table}

\subsection{Targets observed in the SC mode}

\begin{figure*}
\centering
\includegraphics[width=\hsize]{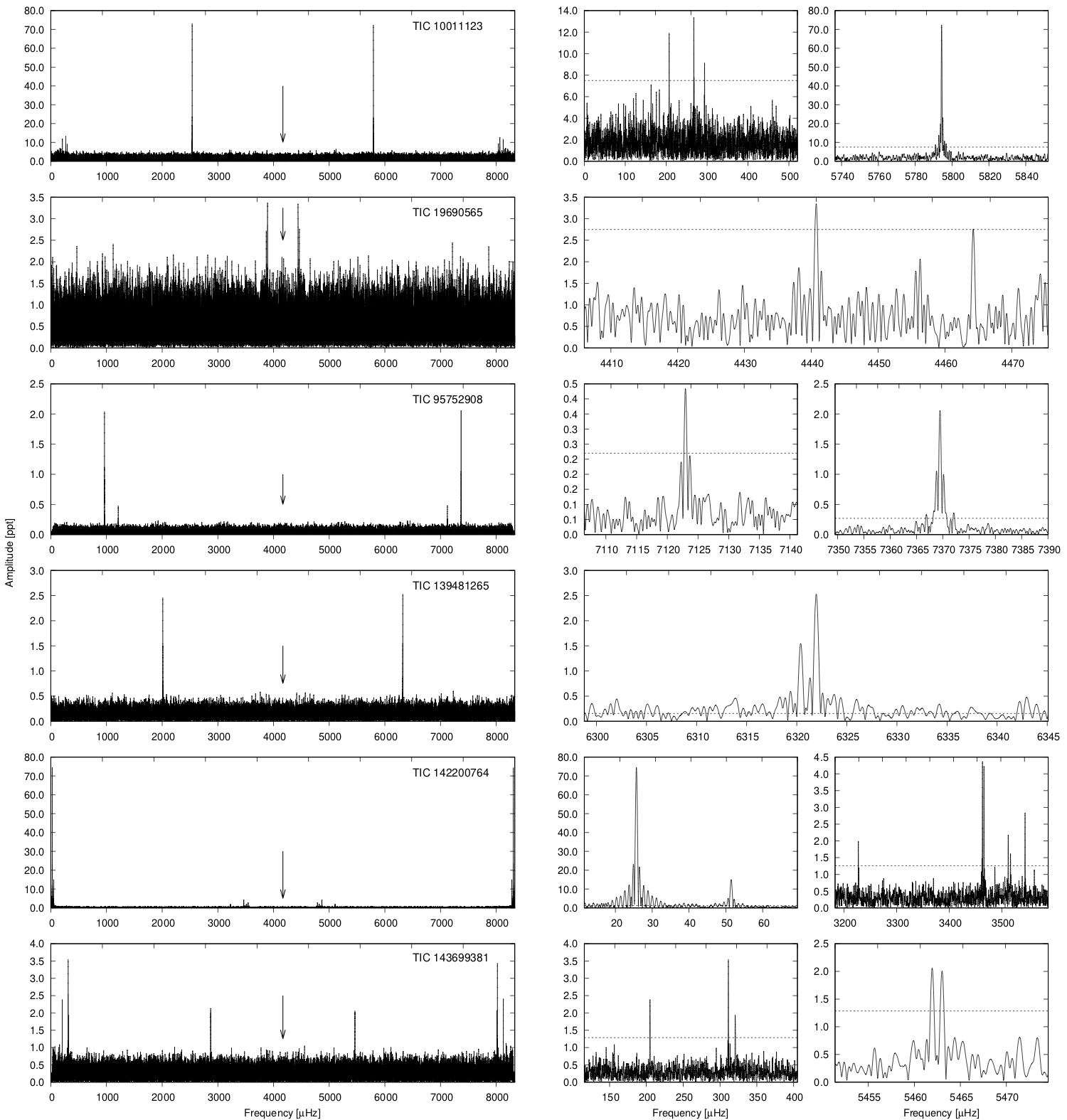}
\caption{Amplitude spectra of the targets observed in the SC mode. {\it Left panels:} Frequency range, up to twice the Nyquist frequency. The arrows point at the Nyquist frequency of 4166.67\,\uHz. {\it Right panels:} Close-ups at the detected frequencies.}
\label{fig:SC_ft1}
\end{figure*}

\paragraph{TIC\,10011123}
(TYC\,4824-1038-1 = Gaia\,DR2\,3058\-814\-547\-877\-917\-056) is a new pulsating sdOB star. The spectral classification is confirmed with a spectrum taken with the 1.9\,m telescope at the South African Astronomical Observatory (SAAO). Details of all classifications obtained by means of SAAO spectra can be found in the paper by Worters et.\,al., which is currently in preparation. \tess\ observed the star during Sector\,33. An amplitude spectrum shown in Fig.\,\ref{fig:SC_ft1} delivers three peaks in the g-mode region and one peak in the high-frequency region. Since the target was observed during just one sector, we are unable to separate a real frequency from its alias across the Nyquist frequency. We show the list of the prewhitened frequencies in Table\,\ref{tab:SCfreq}. Since the majority of p modes detected in pulsating sdB stars are above 4000\,\uHz,\ we arbitrarily prewhitened frequencies in the super-Nyquist region. Based on previous detections of p-mode sdB pulsators, we can see a correlation between the effective temperature and the frequencies of p modes: the lower the temperature, the lower the frequencies at which signals can appear. To verify our frequency choice for targets observed in the SC only, we need either the USC data or at least one sector of additional SC data. The latter argument is clearly seen for TIC\,366656123.

\begin{figure}
\centering
\includegraphics[width=\hsize]{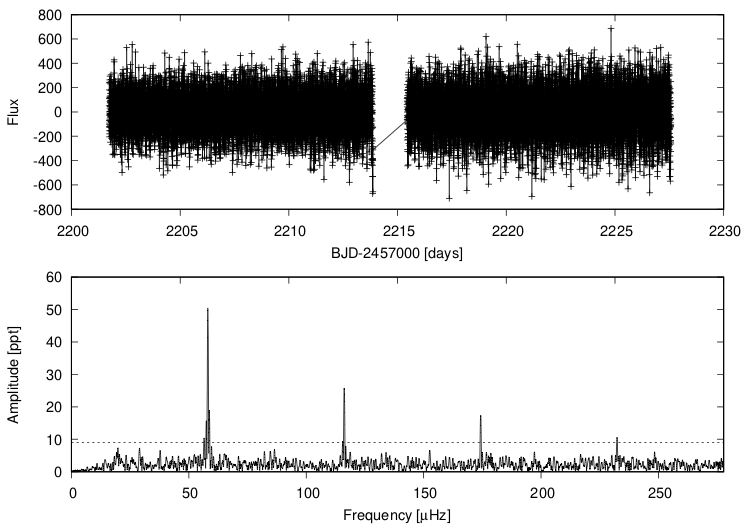}
\caption{\tess\ data of TIC\,754255960. {\it Top panel:} Light curve derived from the Full Frame Images taken in Sector\,33. {\it Bottom panel:} Amplitude spectrum calculated from the light curve sampled at 30\,min. The horizontal line denotes the detection threshold at 4.5 times the median noise level in a residual amplitude spectrum (with two significant peaks removed).}
\label{fig:tic754255960}
\end{figure}

While analysing the contamination of TIC\,1001123, we found TIC\,754255960 to be a variable star. It shows four frequencies in the amplitude spectrum. However, the three higher frequencies all appear to be harmonics of the lower frequency, which indicates non-sinusoidal flux variation. This can be explained, for example, by binarity or pulsations in classical pulsators. Concerning the latter type and given the period of 0.19942(6)\,days, it could be a high-amplitude Delta Scuti, RR\,Lyrae, or anomalous Cepheid star. We retrieved the light curve from the SC data and calculated the amplitude spectrum. Both plots are shown in Fig.\,\ref{fig:tic754255960}.

\paragraph{TIC\,19690565}
(Gaia\,DR2\,5753155495252812544) is a new sdB pulsator. Our fit to a spectrum taken with the 2.56-meter NOT\footnote{All NOT spectra used in this work were obtained with ALFOSC, Grism\,\#18, and a 1.0 arcsec slit, giving R\,=\,1000 and spanning 345-535\,nm.} in 2015 gives \teff\,=\,31\,150(300)\,K, \loggcms\,=\,5.76(5), and $\log$(n(He)/$\log$\,(H))\,<\,-3, indicating an sdB classification. \tess\ observed the star during Sector\,34. We detected two peaks close to the Nyquist frequency and we selected the one in the super Nyquist region. We show the amplitude spectrum in Fig.\,\ref{fig:SC_ft1}, and list the prewhitened frequencies in Table\,\ref{tab:SCfreq}.

\paragraph{TIC\,95752908}
(Gaia\,DR2\,5756471068269831552 = TYC\,4890-19-1) is a known sdB pulsator. \citet{holdsworth17} analyzed the star spectroscopically and classified it as an sdB. The same authors reported up to three frequencies detected in any given amplitude spectrum. However, one frequency, $\upnu_{\rm 3}$, was not detected in all data. \tess\ observed the star during Sector\,8. We detected two frequencies reported by \citet{holdsworth17}. Our frequency f$_{\rm 1}$ is close to $\upnu_{\rm 3}$, while we detected no trace of $\upnu_{\rm 2}$ reported by the authors. We show an amplitude spectrum in Fig.\,\ref{fig:SC_ft1} and list the prewhitened frequencies in Table\,\ref{tab:SCfreq}.

\paragraph{TIC\,139481265}
(Gaia\,DR2\,3342874205845523072) is a new sdB pulsator. Our fit to a spectrum taken with the NOT in 2018 gives \teff\,=\,31\,810(540)\,K, \loggcms\,=\,5.80(4), and $\log$(n(He)/$\log$\,n(H))\,=\,-2.24(8), indicating an sdB classification. \tess\ observed the star during Sector\,33. We found three peaks and we chose those located in the super Nyquist region. The frequencies are listed in Table\,\ref{tab:SCfreq}. They are close to each other, and the separations between the adjacent ones are 1.55(3) and 1.60(5)\,\uHz, respectively, between the two lowest and the two highest frequencies. The frequency spacing is equal within the errors, which indicates that these three peaks can constitute a rotationally split triplet. If so, the rotation period would be 7.35(27)\,d. We show an amplitude spectrum in Fig.\,\ref{fig:SC_ft1}.

\input{SC_table}

\begin{figure}
\centering
\includegraphics[width=\hsize]{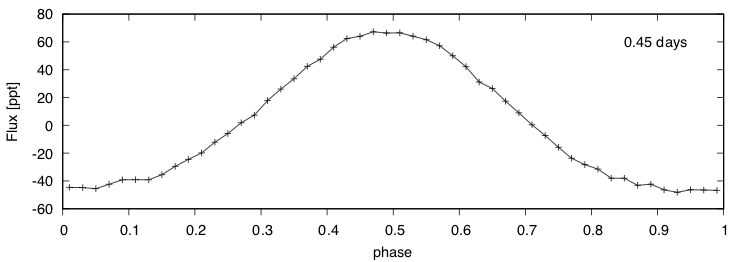}
\caption{Phase-folded light curve of TIC\,142200764 showing a reflection effect. The number in the upper right corner is the orbital period, rounded to two significant digits, used for folding the light curve.
}
\label{fig:tic142200764}
\end{figure}

\paragraph{TIC\,142200764}
(HE\,0230--4323\,=\,Gaia\,DR2\,494702342837980\-2752) was found by \cite{edelmann05} to be a binary system with an sdB as the primary component. The authors reported a binary period of 0.4515(2)\,days, a systemic velocity of 42.3(3)\,\kmsec, and a radial-velocity (RV) amplitude of 109.6(4)\,\kmsec. They concluded that the companion to the sdB star is either an M dwarf or a white dwarf, depending on the orbital inclination and the mass of the sdB. \cite{koen07} confirmed the binarity photometrically and detected pulsations in the sdB star, calling it an unusual pulsating star because the pulsation frequencies changed over the course of several nights from $\sim$32--39\,d$^{-1}$ to $\sim$8--16\,d$^{-1}$. \cite{kilkenny10} revisited the system, explaining the unusual g-mode pulsations as an under-sampling error; with better sampling, they were able to detect short-period pulsations. \tess\ observed the star during Sector\,3. We recovered the orbital frequency and its first harmonic associated with binarity. The phase-folded light curve is shown in Fig.\,\ref{fig:tic142200764}, and the ephemeris is provided in Table\,\ref{tab:ephemerides}. The light curve is dominated by a reflection effect, which points to an M-dwarf as the companion type. The reflection effect can be distinguished from ellipsoidal and Doppler-beaming variations by its high amplitude and the shape of the flux variation. The amplitude of the latter two effects are typically around a few ppt, while reflection can easily reach several percent. Reflection is also easily distinguished from the other effects by the way the first harmonic combines with the orbital period to make the peaks higher, and the troughs shallower, while retaining symmetry around the peak. In addition to the orbital peaks, we found eight frequencies in the p-mode region. We show an amplitude spectrum with two close-ups in Fig.\,\ref{fig:SC_ft1}. Since the star was observed in the SC mode, we would not be able to discern between the real frequencies and their aliases across the Nyquist frequency if the frequencies were not reported by \cite{kilkenny10}. In this case, it is the frequencies in the sub-Nyquist region that are real. Two frequencies, f$_4$ and f$_8$, should be considered tentative since their amplitudes are below our adopted detection threshold, and were also not detected by \cite{kilkenny10}. Frequency f$_5$ was also not detected by the authors. On the other hand, we did not detect frequencies above 5700\,\uHz, which were detected by the authors. The full list of frequencies we detected in the \tess\ data is shown in Table\,\ref{tab:SCfreq}.

\paragraph{TIC\,143699381}
(Gaia\,DR2\,6715490300005795840) is a new sdB pulsator. This spectral classification is obtained with a spectrum taken with the 1.9\,m telescope at SAAO. \tess\ observed the star during Sector\,13. We detected five frequencies: three in the g-mode region and two in the p-mode region, which make the star a hybrid sdB pulsator. We show an amplitude spectrum in Fig.\,\ref{fig:SC_ft1} and list the prewhitened frequencies in Table\,\ref{tab:SCfreq}.

\input{ephemerides}

\begin{figure*}
\centering
\includegraphics[width=\hsize]{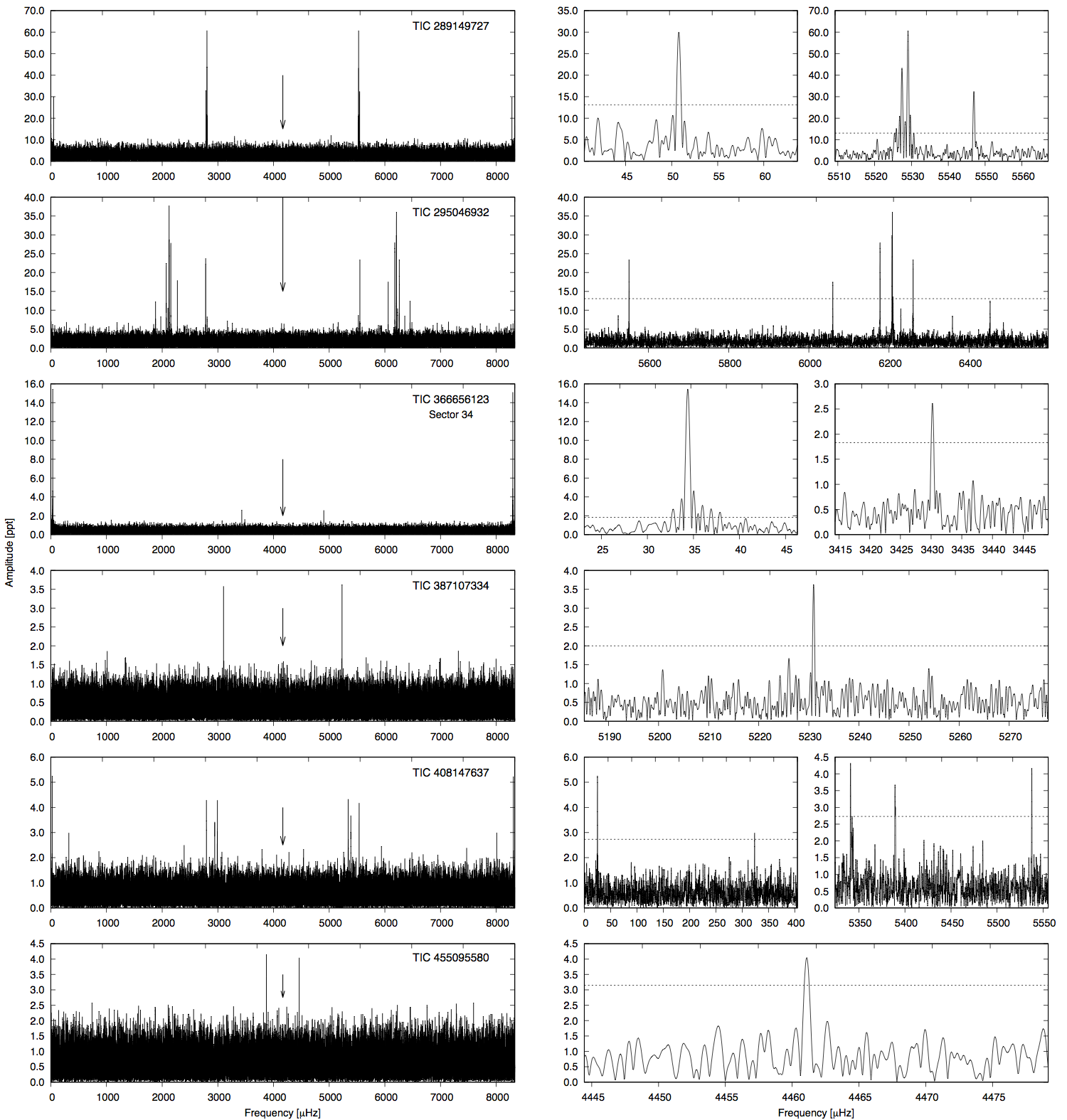}
\caption{Same as in Fig.\,\ref{fig:SC_ft1}, but for another six targets observed only in the SC mode.}
\label{fig:SC_ft2}
\end{figure*}

\paragraph{TIC\,289149727}
(Gaia\,DR2\,5870233314477487872) is a new sdB pulsator. Based on a spectrum taken with the 1.9\,m telescope at SAAO, we classified it as an sdB star. \tess\ observed the star during Sector\,38. We detected four frequencies: one low frequency and three frequencies in the p-mode region. The low frequency is not typical of g-mode pulsations in hot-subdwarf stars and it is probably a signature of binarity rather than a g-mode pulsation. As this target is rather faint and residing in the crowded environment of the galactic plane, the light curve is too noisy to distinguish the harmonic of the suspected orbital peak. We show an amplitude spectrum in Fig.\,\ref{fig:SC_ft2} and list the prewhitened frequencies in Table\,\ref{tab:SCfreq}.

\paragraph{TIC\,295046932}
(Gaia\,DR2\,5811947687666193280) is a new sdB pulsator. We classified the star as an sdB based on a spectrum taken with the 1.9 m telescope at SAAO. \tess\ observed the star during Sector\,39. We detected 11 frequencies in the p-mode region. We show an amplitude spectrum in Fig.\,\ref{fig:SC_ft2} and list the prewhitened frequencies in Table\,\ref{tab:SCfreq}.

\begin{figure}
\centering
\includegraphics[width=\hsize]{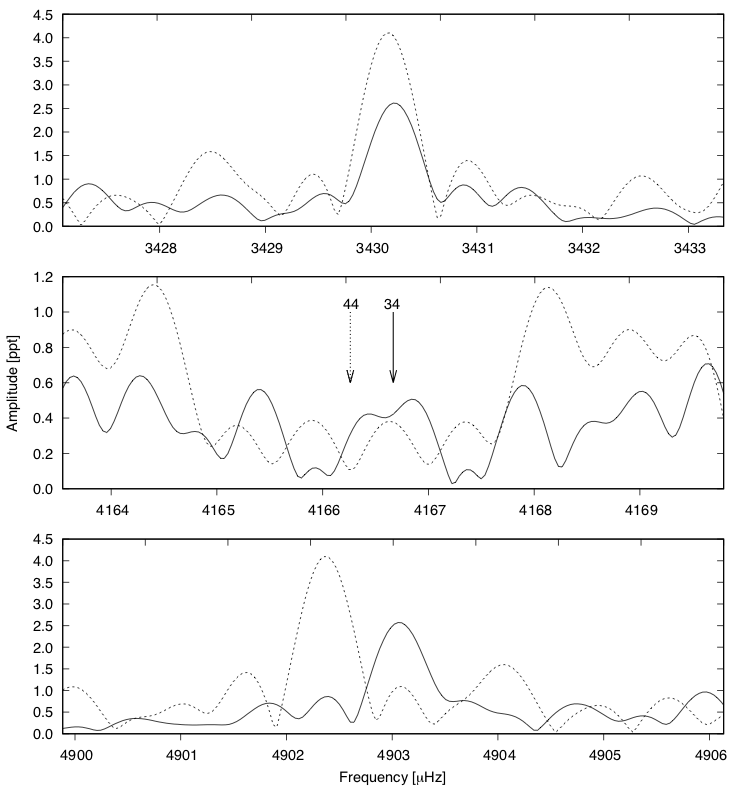}
\caption{Amplitude spectrum of TIC\,366656123. {\it Top panel:} Close-up of the amplitude spectrum showing a high frequency in the sub-Nyquist region. Solid and dashed lines represent Sectors\,34 and 44, respectively. {\it Middle panel:} Nyquist-frequency region. The arrows point at the Nyquist frequency in Sectors\,34 and 44. {\it Bottom panel:} Close-up of the amplitude spectrum showing a high frequency in the super-Nyquist region.}
\label{fig:tic366656123}
\end{figure}

\paragraph{TIC\,366656123}
(Gaia\,DR2\,595128265015393152 = SDSS J084122.66+063029.6) is a new sdB pulsator. Our fit to a spectrum taken with the 2.5 m Sloan Digital Sky Survey (SDSS) telescope in 2008 gives \teff\,=\,31\,300(200)\,K, \loggcms\,=\,5.56(3), and $\log$(n(He)/\,n(H))\,=\,-2.8(1). It was included in a sample of hot-subdwarf candidates analyzed by \citet{sahoo20a}, who detected a low frequency, and consequently the object was proposed and successfully observed by \tess\ during Sectors\,34 and 44. The data are too far apart to combine and analyze as one piece since the Fourier window function is very complex. Instead, we analyzed these two data sets separately and compared the signals detected. In both sectors we detected one frequency in the g-mode region and one high frequency in the p-mode region. We show an amplitude spectrum calculated only from Sector\,34 data in Fig.\,\ref{fig:SC_ft2}. To be consistent with analyses of other targets, we should pick the frequency in the super Nyquist region. However, the high frequency is shifted between sectors (shown in Fig.\,\ref{fig:tic366656123}). The middle panel shows that the Nyquist frequency is shifted between sectors, which will consequently shift the reflections of frequencies. Indeed, the frequencies in the super-Nyquist region are not aligned, which we interpret as reflections, and therefore the signal at high frequency originates in the sub-Nyquist region. This also confirms that our arbitrary choice of selecting peaks in the super-Nyquist region may not always be correct. We list the prewhitened frequencies in Table\,\ref{tab:SCfreq}.

\paragraph{TIC\,387107334}
(Gaia\,DR2\,6439287618985307776 = BPS CS 22959-140) is a new sdB pulsator. It was originally classified as an sdO subdwarf in the BPS catalog \citep{Beers92}. However, a spectrum taken with the 1.9\,m telescope at SAAO indicates an sdB classification. \tess\ observed the star during Sector\,13. We detected only a single frequency in the p-mode region. We show an amplitude spectrum in Fig.\,\ref{fig:SC_ft2} and list the prewhitened frequency in Table\,\ref{tab:SCfreq}.

\paragraph{TIC\,408147637}
(Gaia\,DR2\,5867781918951429760) is a new sdB pulsator. Based on a spectrum taken with the 1.9 m telescope at SAAO, we classified it as an sdB star. \tess\ observed the star during Sector\,38. We detected five frequencies: one low frequency, one in the g-mode region, and three in the p-mode region. We show an amplitude spectrum in Fig.\,\ref{fig:SC_ft2} and list the prewhitened frequencies in Table\,\ref{tab:SCfreq}.

\paragraph{TIC\,455095580}
(Gaia\,DR2\,3096564462848659328) is a new sdB pulsator. Our fit to a spectrum taken with the NOT in 2022 gives \teff\,=\,30\,800(500)\,K, \loggcms\,=\,5.7(1), and $\log$(n(He)/$\log$\,(H))\,=\,-2.55(15), indicating an sdB classification. \tess\ observed the star in Sector\,34. We detected only one frequency in the p-mode region. We show an amplitude spectrum in Fig.\,\ref{fig:SC_ft2} and we list the prewhitened frequency in Table\,\ref{tab:SCfreq}.

\subsection{Targets observed in the USC mode}
For all targets observed in the USC mode, there is also corresponding SC data available. In addition, some targets listed below were observed during specific sectors (1--13) in the SC mode only. Since the USC data provide us with a unique frequency identification, we decided not to include any SC data, even those taken in sectors without available USC data. Table\,\ref{tab:targets_all} provides detailed sector information and we do not mention sectors with the SC data below.

\begin{table}{}
\caption{List of frequencies we detected in the targets observed in the USC mode. The table is available online.}
\label{tab:USCfreq}
\end{table}

\begin{figure*}
\centering
\includegraphics[width=\hsize]{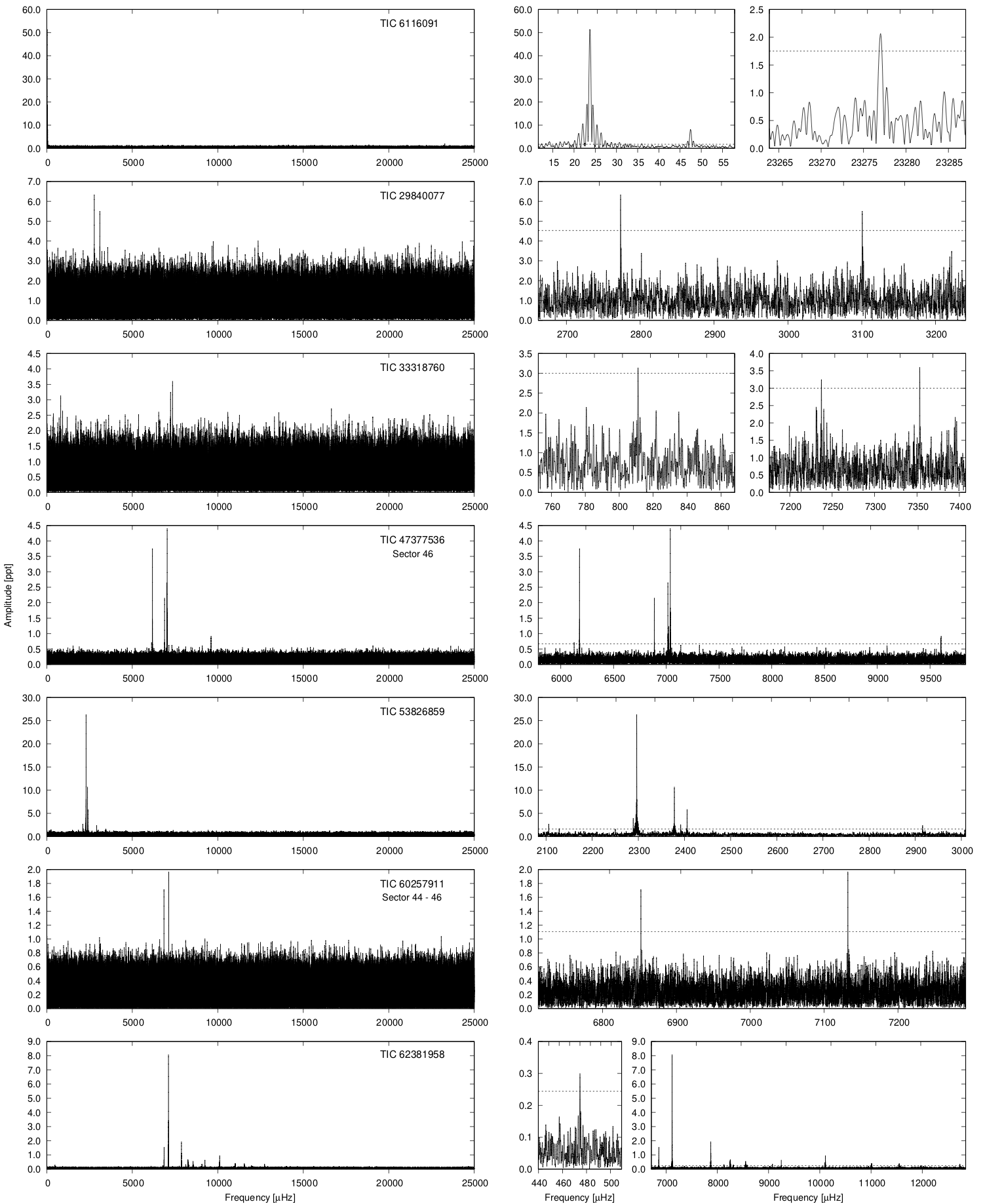}
\caption{Amplitude spectra of targets observed in the USC mode. {\it Left panels:} Frequency range up to the Nyquist frequency. {\it Right panels:} Close-ups at the detected frequencies.}
\label{fig:USC_ft1}
\end{figure*}

\paragraph{TIC\,6116091}
(Gaia\,DR2\,6196248648201755904, EC\,13185--2111, HE\,1318--2111) is a new sdOB pulsator. It is included in the Edinburgh\,--\,Cape (EC) survey, where it is classified as an sdB star \citep{kilkenny97b}. Later, \citet{christlieb01} rediscovered it in the Hamburg/ESO survey, and classified it as an sdOB star. \citet{stroeer07} analyzed an Ultraviolet and Visual Echelle Spectrograph (UVES) spectrum from the SPY survey, and found \teff\,=\,36\,254\,K, \loggcms\,=\,5.42, and $\log$(n(He)/$\log$\,(H))\,=\,-2.91. The target is listed with a period of 0.487502\,d and an RV amplitude of 48.5(1.2)\,\kmsec\ in a list of close binaries by \citet{geier11a}, citing an unpublished Napiwotzki et al.\,(in preparation) paper \citet{sahoo20a} used Full Frame Images taken during Sector\,10 and reported a 0.49\,d variation. Consequently the object was proposed and successfully observed by \tess\ during Sector\,37. We confirmed the low frequency, and we also detected its harmonic. Based on the shape and the high amplitude of the flux variation, the orbital signal must be a reflection effect, as in the case of TIC\,142200764. We detected only one significant high frequency. We show an amplitude spectrum in Fig.\,\ref{fig:USC_ft1} and list the prewhitened high frequency in Table\,\ref{tab:USCfreq}.

\paragraph{TIC\,29840077}
(Gaia\,DR2\,6803823552347267968, EC\,21032--2551) is a new sdB pulsator. This target was identified as an sdB star by \citet{odonoghue13}. \tess\ observed the star during Sector\,28. We detected two frequencies in the p-mode region. We show an amplitude spectrum in Fig.\,\ref{fig:USC_ft1} and list the prewhitened frequencies in Table\,\ref{tab:USCfreq}.

\begin{figure*}
\centering
\includegraphics[width=\hsize]{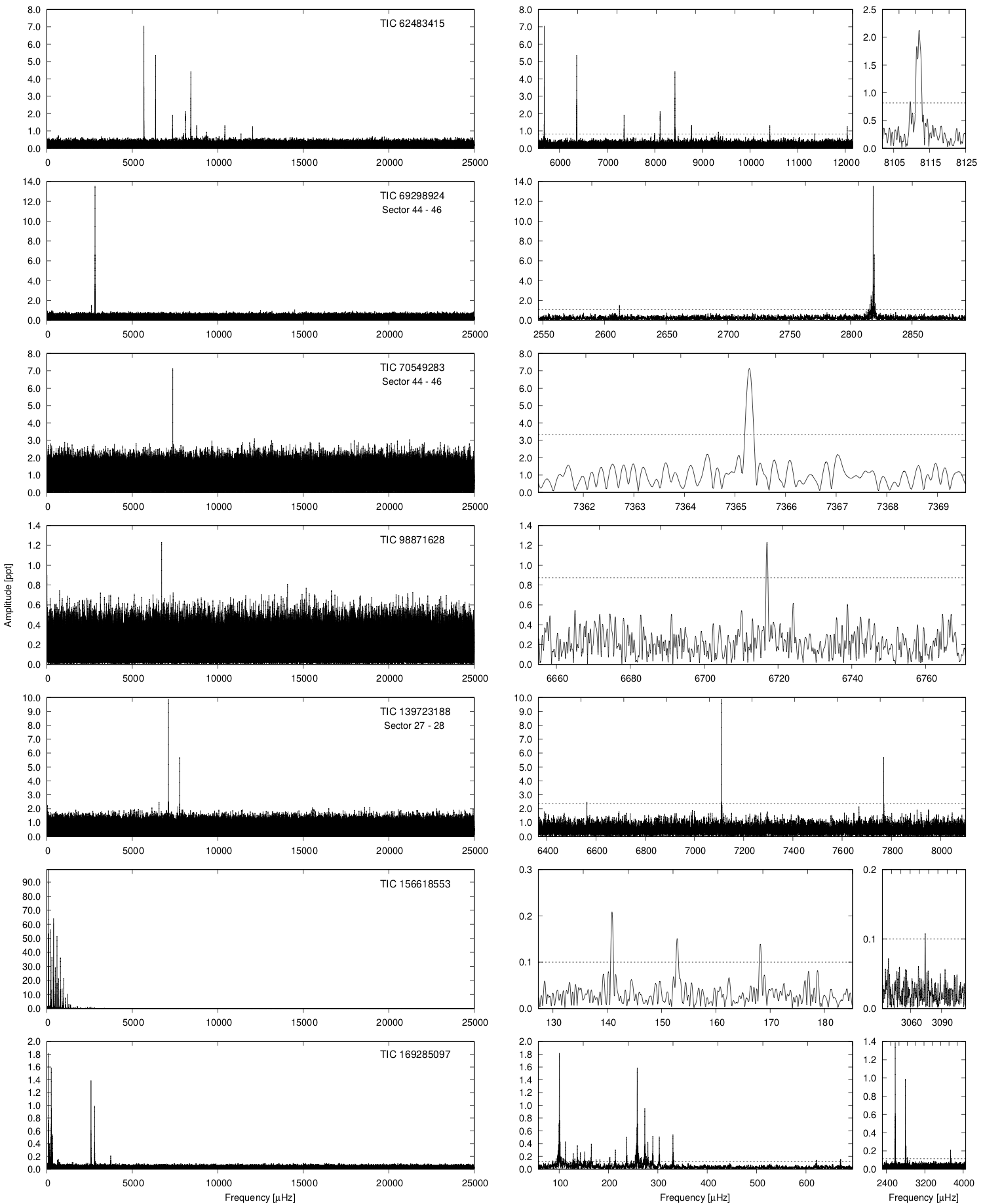}
\caption{Same as in Fig.\,\ref{fig:USC_ft1}, but for another seven targets observed in the USC mode.}
\label{fig:USC_ft2}
\end{figure*}

\begin{figure*}
\centering
\includegraphics[width=\hsize]{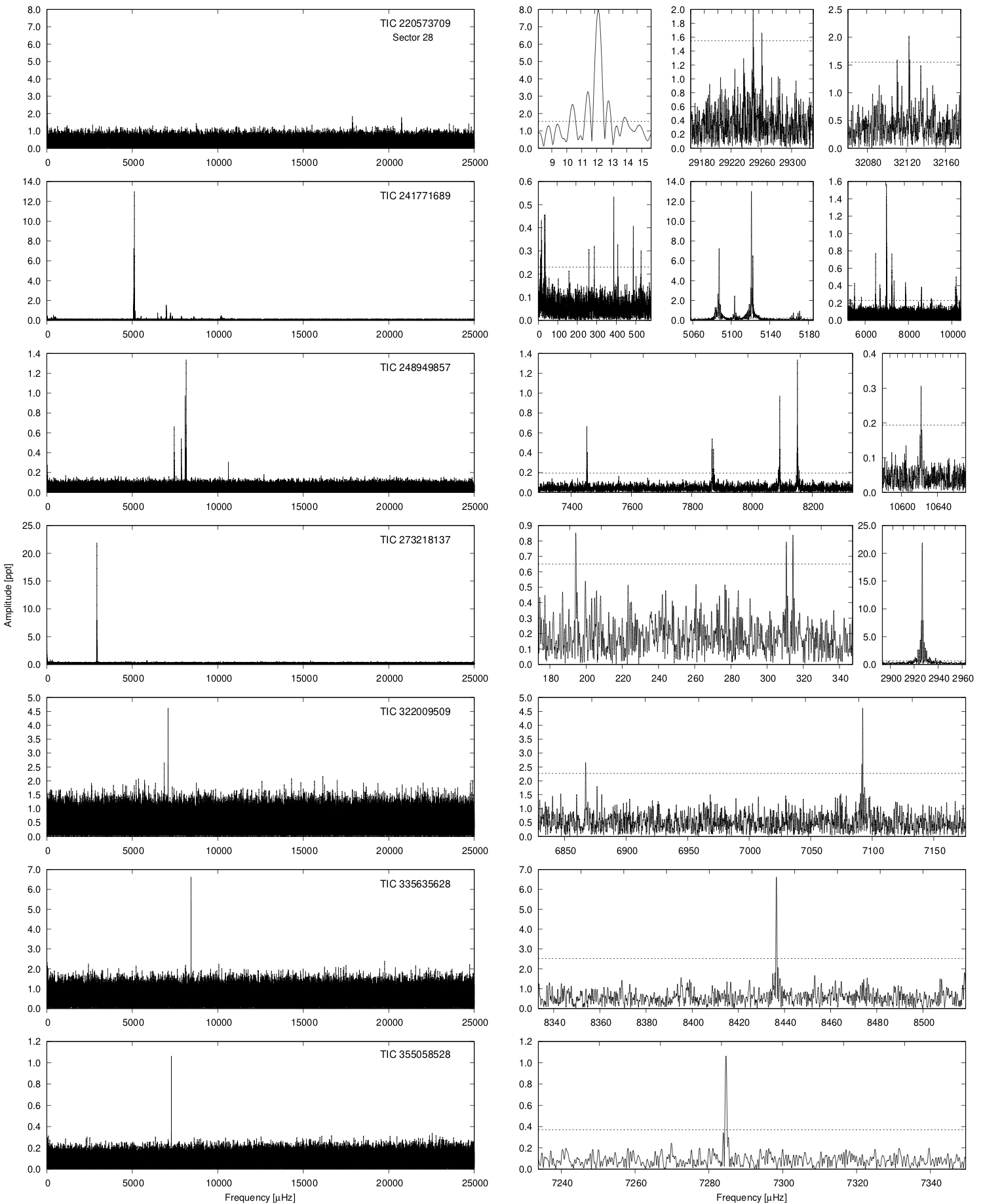}
\caption{Same as in Fig.\,\ref{fig:USC_ft1} but for another seven targets observed in the USC.}
\label{fig:USC_ft3}
\end{figure*}

\paragraph{TIC\,33318760}
(Gaia\,DR2\,3766481985523752320, V541\,Hya, EC\,09582--1137, PG\,0958--116) is a known sdB pulsator. It was included in the catalog of UV-excess stellar objects \citep{green86}, in the catalog of spectroscopically identified hot subdwarfs \citep{kilkenny88}, and in the EC survey \citep{kilkenny97b}. \citet{kilkenny06} analyzed photometric data and reported a detection of two frequencies in the p-mode region. Additionally, four significant frequencies were reported by \citet{randall09}. One of the modes was considered split by rotation. The authors performed period fits and obtained structural parameters of the sdB star. \citet{mackebrandt20} searched, with a null result, for a change in the arrival time of stellar pulsations induced by a sub-stellar companion. \tess\ observed the star during Sector\,35. We detected one frequency (f$_{\rm 1}$) in the intermediate-frequency region and two frequencies (f$_{\rm 2}$, f$_{\rm 3}$) in the p-mode region. Frequencies f$_{\rm 1}$ and f$_{\rm 2}$ are of low amplitude and could be noise-induced. These two frequencies had not been detected from the ground. Frequency f$_{\rm 3}$ confirms the highest frequency reported by both \citet{kilkenny06} and \citet{randall09}. Its amplitude is much smaller in the \tess\ data than in the ground observations. The intrinsic amplitude could have changed since the last time the star was monitored, or remained the same but the wide band pass of the \tess\ photometric system could have averaged the flux variation. While the intrinsic effect is not easy to predict, it certainly takes place. We show an amplitude spectrum in Fig.\,\ref{fig:USC_ft1} and list the prewhitened frequencies in Table\,\ref{tab:USCfreq}.

\paragraph{TIC\,47377536}
(Gaia\,DR2\,3806303066866089216, UY\,Sex, PG\,1047+003) is a known sdB pulsator and has been extensively studied since its independent discovery by \citet{billeres97} and \citet{odonoghue98}. Five and eight independent frequencies were reported by the former and latter authors, respectively. \citet{odonoghue98} confirmed all frequencies reported by \citet{billeres97}. \citet{kilkenny02} reported an enhanced list of frequencies obtained from a multisite campaign. The authors listed 18 frequencies, but three reported by \citet{odonoghue98} were not detected. \citet{kilkenny02} speculated on rotationally split modes, and provided a mode identification and a comparison of modeled and observed periods. \citet{charpinet03} published a model providing mode identifications for 16 frequencies. The star was observed during the \kep\ \ktwo\ mission \citep{reed20}. The authors detected 97 frequencies, including rotationally split multiplets. The frequency splitting allowed a rotation period of 24.6(3.5)\,days to be estimated. The authors reported all frequencies detected prior to their analysis. \tess\ observed the star in Sectors\,35 and 46. We analyzed both sectors separately. Since we did not detect all frequencies in both sectors, while the amplitudes of those repeating in both sectors are different, we decided not to merge data for a joint analysis. Merging would average amplitudes and contribute to a complex window function, making prewhitening difficult. In both sectors we detected only 13 independent frequencies. The majority of these frequencies confirm the ones reported by \citet{reed20}. However, even allowing for a poor frequency resolution, around 0.5\,\uHz, we report f$_{\rm 7}$ in Sector\,35, and f$_{\rm 4}$ and f$_{\rm 5}$ in Sector\,46 as new detections. We show an amplitude spectrum calculated only from Sector\,46 data in Fig.\,\ref{fig:USC_ft1} and list the prewhitened frequencies in Table\,\ref{tab:USCfreq}.

\paragraph{TIC\,53826859}
(Gaia\,DR2\,2921084812241684608) is a new sdB pulsator. The sdB classification is confirmed with a spectrum taken with the 1.9 m telescope at SAAO. \tess\ observed the star during Sector\,33 and we detected 11 frequencies in the p-mode region, although the frequencies are rather low for p-modes. We show an amplitude spectrum in Fig.\,\ref{fig:USC_ft1} and list the prewhitened frequencies in Table\,\ref{tab:USCfreq}. 

\paragraph{TIC\,60257911}
(Gaia\,DR2\,656345533398638464, EPIC\,211\-823\-779) is a known sdB pulsator. \citet{reed18a} found the star to be a binary consisting of an sdB star and a main-sequence companion. The authors analyzed \ktwo\ data and reported 16 frequencies in the p-mode region. Just a few candidates for rotationally split modes were marked and a possible 11.5(8)-day spin rate was estimated. \tess\ observed the star during Sectors\,44, 45, and 46. We detected only two frequencies, which were found in the \ktwo\ data to be the highest-amplitude frequencies. We do not confirm our frequency f$_{\rm 2}$ to be rotationally split. The \tess\ data coverage of $\sim$79\,d is comparable to the coverage obtained with the \ktwo\ data. We show an amplitude spectrum in Fig.\,\ref{fig:USC_ft1} and list the prewhitened frequencies in Table\,\ref{tab:USCfreq}.

\paragraph{TIC\,62381958}
(Gaia\,DR2\,5149241067178231552, EC\,01541-1409) is a known sdOB pulsator. It was noted as an sdOB star in the EC survey \citep[as published in][]{kilkenny16a}, and \citet{kilkenny09} analyzed photometric data and reported six frequencies in the p-mode region. Follow-up multisite time-series data were collected and analyzed by \citet{reed12a}. The number of frequencies reported depends on one observing season but reaches almost 30 in 2009. All frequencies reported by \citet{kilkenny09} were detected by \citet{reed12a}. In the latter paper, the time span of the 2009 campaign was around a month, yet still the authors reported no candidate rotationally split frequencies. \citet{randall14} reported results of a mode identification of several frequencies detected in the spectrophotometric data collected by the authors. \tess\ observed the star during Sector\,30. We show an amplitude spectrum in Fig.\,\ref{fig:USC_ft1} and list the prewhitened frequencies in Table\,\ref{tab:USCfreq}. The majority of frequencies we detected overlap with those listed by \citet{reed12a}. In addition, we found one significant frequency in the g-mode region at a relatively high frequency. Such a frequency in this star has not been reported thus far. As shown by \citet{reed12a}, this star exhibits frequency and/or amplitude variation of the pulsations modes between two observational seasons. The signal in the amplitude spectrum calculated from the \tess\ data is also very unstable. We tried to prewhiten all signals down to the adopted threshold, but we do not know if the close frequencies, within 1\,\uHz, are independent or a consequence of an instability of one mode. A common frequency spacing between close frequencies is about 0.5\,\uHz, which is comparable to the frequency resolution  derived from 1/T, with T being the total data coverage. Since the data are quite continuous, the aliases in a Fourier window response are of low amplitude; hence, we do not expect the close frequencies to be aliases.

\paragraph{TIC\,62483415}
(Gaia\,DR2\,6601695863046409600, PHL\,252, EC\,22221-3152) is a known sdOB pulsator \citep{kilkenny16a}. \citet{kilkenny09} observed it photometrically and reported 11 frequencies between 5\,500 and 11\,900\,\uHz. It was not mentioned in that paper, but f$_{\rm 11}$ would appear to be the first harmonic of f$_{\rm 1}$. \citet{barlow17} delivered two-site photometry of the star and reported 11 independent frequencies, along with three combination and four close frequencies, which the authors interpreted as rotationally split modes. The rotation period was estimated at $\sim$8\,d but this is incorrect (see below). \tess\ observed the star during Sector\,28. We detected ten independent frequencies, as well as two combination frequencies. We show an amplitude spectrum in Fig.\,\ref{fig:USC_ft2} and list the prewhitened frequencies in Table\,\ref{tab:USCfreq}. We found f$_{\rm 11}$ and f$_{\rm 12}$, which were not detected by the other authors. The rotation period derived by \citet{barlow17} is close to their data coverage and that is why they suggested a longer timebase to confirm this result. One sector of \tess\ data is roughly 27\,d. If the rotation period is indeed about 8\,d, we should surely detect split modes. They listed their f$_{\rm 1}$ and f$_{\rm 5}$ as candidates for split modes. We find the former frequency (f$_{\rm 5}$ in our list) to be an amplitude variable, and the latter frequency (f$_{\rm 6}$ in our list) to be single. Frequency f$_{\rm 5}$ in our list is so variable that we were unable to prewhiten any frequency in that range and we decided to quote numbers read by eye; hence, these are given without error estimates.

\paragraph{TIC\,69298924}
(Gaia\,DR2\,654866823401111168, GALEX J080656.7+152718) is a known sdB pulsator. \citet{vennes11} identified it as an sdB star, while \citet{baran11a} analyzed photometric data reporting four frequencies, two in the g-mode region and two in the p-mode region, showing the star to be a hybrid sdB pulsator. The authors analyzed time-series spectroscopic data, detecting one frequency, which overlaps with the highest-amplitude frequency in the photometric data. \tess\ observed the star during Sectors\,44, 45, and 46. Since these are consecutive sectors and the amplitude spectra calculated from single-sector data do not differ substantially, we combined all the data. We detected four frequencies in the p-mode region and none in the g-mode region. Three frequencies are close to each other and seem to be rotationally split frequencies. The average frequency splitting of 0.8953(67)\,\uHz\ translates into a rotation period of 12.9(1)\,d. These three frequencies overlap with, apparently unresolved, frequency f${\rm 1}$ reported by \citet{baran11a}. Our frequency f$_{\rm 1}$ is close to the frequency f$_{\rm 1}$ reported earlier. We show an amplitude spectrum in Fig.\,\ref{fig:USC_ft2} and list the prewhitened frequencies in Table\,\ref{tab:USCfreq}.

\paragraph{TIC\,70549283}
(Gaia\,DR2\,673058556816796288) is a new sdOB pulsator. The spectral classification was reported by \citet{lei19}. \tess\ observed the star during Sectors\,44, 45\ and 46. We detected only one frequency in the two-sector merged data. We show an amplitude spectrum in Fig.\,\ref{fig:USC_ft2} and list the prewhitened frequency in Table\,\ref{tab:USCfreq}.

\paragraph{TIC\,98871628}
(Gaia\,DR2\,3486707300467202304, V551\,Hya, EC\,11583-2708) is a known sdB pulsator. \citet{kilkenny97b} listed the star as having a composite spectrum (sdB+). Pulsations were found by \citet{kilkenny06}, who detected four frequencies and noted that the companion could be an early G-type star. \tess\ observed the star during Sector\,36. We detected only one frequency, which is the one with the highest amplitude reported by \citet{kilkenny06}. We show an amplitude spectrum in Fig.\,\ref{fig:USC_ft2} and list the prewhitened frequency in Table\,\ref{tab:USCfreq}.

\paragraph{TIC\,139723188}
(Gaia\,DR2\,6466786576593357440, EC\,21281-5010) is a known sdOB pulsator. \citet{kilkenny15} identified the star as a hot subdwarf, and \citet{kilkenny19} observed it photometrically and reported three frequencies. \tess\ observed the star during Sectors\,27 and 28. Since these are consecutive sectors and the amplitude spectra calculated from single sectors show very similar signal distribution, we analyzed these two sectors merged. We detected three frequencies, but only two are the same as those reported by \citet{kilkenny19}. We show an amplitude spectrum in Fig.\,\ref{fig:USC_ft2} and list the prewhitened frequencies in Table\,\ref{tab:USCfreq}.

\paragraph{TIC\,156618553}
(Gaia\,DR2\,3675067076961979264, HW\,Vir, PG\,1241-084, HE\,1241-0823) is a known sdB pulsator. It has been extensively studied in the past. \citet{berger80} classified the object as an sdB star, while \citet{menzies86} found it to be an eclipsing binary system. Since that time the object has been monitored mostly photometrically to verify the stability of the orbital period. Variation in the orbital period was first detected by \citet{kilkenny94} and has been extensively monitored since then, with a recent compilation by \citet{baran18}. We estimated the mid-times of eclipses observed by \tess\ and derived the orbital period of 0.116719509(8)\,d. \citet{baran18} found the sdB star to be a rich pulsator, reporting both the g- and p-mode pulsations. In total, they listed 91 frequencies, of which the majority are in the g-mode region, with a decent number in the intermediate region and only three above 3000\,\uHz. The authors delivered the most updated Observed--Calculated (O--C) diagram at the time. \tess\ observed the star during Sector\,46. We removed the orbital contribution in the amplitude spectrum prior to pulsation search. Finally, we detected only four frequencies, with one being in the p-mode region. Only one frequency f$_{\rm 2}$ we detected in the \tess\ data overlaps with f$_{\rm 4}$ listed by \citet{baran18}. We show an amplitude spectrum in Fig.\,\ref{fig:USC_ft2} and list the prewhitened frequencies in Table\,\ref{tab:USCfreq}. We also removed the pulsations from the data to derive the mid-times of the eclipses. We folded all eclipses within a single \tess\ orbit and derived the mid-times 2\,459\,559.135104(1) and 2\,459\,572.908007(1) during the first and the second \tess\ orbits, respectively.

\paragraph{TIC\,169285097}
(Gaia\,DR2\,2312392250224668288, HE\,2341-3443, CD-35\,15910) is a new sdB pulsator. It was listed as an sdB star by \citet{lamontagne00}. During the \tess\ mission, it was first observed during Sector\,2 and the results of the pulsation analysis were reported by \citet{sahoo20b}. The authors analyzed only the SC data reporting 43 frequencies, with six being high frequencies. \tess\ photometry was also taken during Sector\,29. We list the USC frequencies in Table\,\ref{tab:USCfreq} and show an amplitude spectrum in Fig.\,\ref{fig:USC_ft2}. We detected only 34 frequencies, which is nine fewer than in the list reported by \citet{sahoo20b}, and we did not find any new frequencies in Sector\,29.

\paragraph{TIC\,220573709}
(Gaia\,DR2\,4720417758386878080, EC\,03089-6421) is a known sdO pulsator. \citet{kilkenny15} listed it as an sdO star, and \citet{kilkenny17} reported two very high frequencies, at $\sim$32.12\,mHz and $\sim$29.26\,mHz, detected in the photometric data. The authors speculated that this star could be a field counterpart of the $\upomega$\,Cen sdO variables \citep{randall11}. Later, \citet{kilkenny19} reobserved the star, finding an additional frequency at $\sim$37.65\,mHz. \tess\ observed the star during Sectors\,28 and 30. The USC data have a Nyquist frequency shorter than the frequencies reported by \citet{kilkenny17,kilkenny19} and so, if we did not have the ground-based results, it would be difficult to extract the correct frequencies from the aliases. The amplitude spectra from the two sectors look a little different, so the window function of the combined data gives averaged amplitudes and split peaks, which make prewhitening more difficult. Therefore, we decided to analyze each sector data separately. We show an amplitude spectrum calculated only from Sector\,28 data in Fig.\,\ref{fig:USC_ft3} and list the prewhitened frequencies in Table\,\ref{tab:USCfreq}. In both sectors we confirm frequencies around 32.12\,mHz, while we detected the 29.26\,mHz frequency only in Sector\,28. We found no signature of the highest frequency around 37.65\,mHz. Besides the high frequencies, in both sectors, we detect $\sim$12.1\,\uHz\ frequency. Such a low frequency can be attributed either to a binary frequency or to on-board systematics. We know of a 1\,d$^{-1}$ artifact, which is related to the Earth's rotation and is explained in the \tess\ Data Release Notes available on the MAST. However, two high frequencies are spaced exactly by the low-frequency value (see Table\,\ref{tab:USCfreq}). To interpret both the low frequency and the spacing between high frequencies, we can invoke a tidally locked binary system. In this case, the rotation, determined by the frequency splitting between rotationally split frequencies, would be the same as the binary frequency. If this interpretation is correct, the star has the shortest rotation period among all subdwarfs for which the rotation has been estimated. We stress that not all high frequencies are split, which can weaken our interpretation, though the side components are independent from the central one, and their absence does not necessarily rule out our conclusion.

\paragraph{TIC\,241771689}
(Gaia\,DR2\,6093621087563287040, CD-48\,8608) is a new sdB pulsator. It is included in the GALEX survey \citep{nemeth12}. \citet{kawka15} listed the star as a binary system consisting of an sdB star and a G8 main-sequence companion. \tess\ observed the star during Sector\,38. We show an amplitude spectrum in Fig.\,\ref{fig:USC_ft3}. It is very rich in frequencies in the p-mode region. We also detected six frequencies in the g-mode region, so the star is a hybrid pulsator. The list of prewhitened frequencies is shown in Table\,\ref{tab:USCfreq}. The highest-amplitude frequencies $\sim$5120\,\uHz\ show a multiplet structure and there are two frequency splittings that we can report. The splitting between frequencies f$_{\rm 9}$, f$_{\rm 10}$, and f$_{\rm 11}$, between f$_{\rm 12}$ and f$_{\rm 13}$, and between f$_{\rm 15}$ and f$_{\rm 16}$ is $\sim$1.5\,\uHz. We detected a few low-amplitude frequencies between 10 and 40\,\uHz, including 15.7\,\uHz\ and its first harmonic, and 17.4\,\uHz\ and its first harmonic. The frequencies are not included in our list of prewhitened frequencies. If the splitting of $\sim$1.5\,\uHz\ is caused by rotation, then the period would be $\sim$7.7\,d, which is comparable, for example, to that estimated for V585\,Peg \citep{baran09}. Some of the frequencies above 10\,000\,\uHz\ can be combination frequencies of the highest-amplitude modes. We found no exact values, but f$_{\rm 42}$ is very close to 2$\cdot$f$_{\rm 15}$, while f$_{\rm 41}$ is close to f$_{\rm 10}$+f$_{\rm 15}$.

\paragraph{TIC\,248949857}
(Gaia\,DR2\,2482171590176492928, EO\,Cet, PB\,8783) is a known sdO pulsator. It had been originally identified as an sdB star in which pulsations were found by \citet{koen97}, who reported six high frequencies. A subsequent paper by \cite{odonoghue98} reported 11 frequencies, including resolved close frequencies, as a consequence of a higher-frequency resolution, but did not confirm the 7.883 and 8.291\,mHz frequencies. The authors provided a mode identification by fitting models to the observed periods. \citet{ostensen12a} analyzed new spectra and reclassified this star as an sdO star. \tess\ observed the star during Sector\,30. We detected 11 frequencies, confirming detections reported by \citet{odonoghue98} and \citet{vangrootel14}. We show an amplitude spectrum in Fig.\,\ref{fig:USC_ft3} and list the prewhitened frequencies in Table\,\ref{tab:USCfreq}.

\paragraph{TIC\,273218137}
(Gaia\,DR2\,5371215147518355328, L\,325-214) is a new sdB pulsator. It was mistakenly identified as a DA white dwarf by \citet{mccook87}. \citet{kawka07} collected new spectroscopy and reclassified the object as an sdB star, with \teff\,=\,30\,080(660)\,K, \loggcms\,=\,5.15(10), and $\log$(n(He)/$\log$\,(H))\,$\lesssim$\,--3.0. \tess\ observed the star during Sector\,37. We detected a large-amplitude high frequency with two low-amplitude frequencies symmetrically spaced by 3.696(21)\,\uHz, on average. We show an amplitude spectrum in Fig.\,\ref{fig:USC_ft3} and list the prewhitened frequencies in Table\,\ref{tab:USCfreq}. We calculated the O--C diagram from the phase of frequency f$_{\rm 5}$ and  detected a 3.15\,d variation. Independently of our analysis, \citet{smith22} reported the detection of a secondary companion with a period of 3.1\,d.

\paragraph{TIC\,322009509}
(Gaia\,DR2\,4990641054653089664, JL\,166) is a known sdOB pulsator. The \citet{kilkenny88} catalog lists it as an sdOB star and \citet{barlow09} reported ten frequencies and one combination frequency. \tess\ observed the star in Sector\,29. We show an amplitude spectrum in Fig.\,\ref{fig:USC_ft3} and list the prewhitened frequencies in Table\,\ref{tab:USCfreq}. We detected only two frequencies and neither were previously reported by \cite{barlow09}, although they are close to their $f_2$ and $f_4$.

\paragraph{TIC\,335635628}
(Gaia\,DR2\,3609593392911348096, PG\,1315-123) is a known sdB pulsator. \citet{green86} classified the star as a cataclysmic variable, while \citet{kilkenny88} listed it as an sdB star. \citet{reed19} analyzed data collected during the \ktwo\ mission and reported 46 frequencies in the p-mode region and 16 frequencies in the g-mode region. The g-modes are rather surprising at the high \teff\ of $\sim$36\,000\,K. Among both p- and g-modes, the authors identified consistent rotationally split frequencies, derived the rotation period of 16.18(57)\,d, and concluded that the star rotates as if it were a solid body. The authors analyzed spectroscopic data, reporting a clear contribution from a main-sequence companion. \tess\ observed the star during Sector\,46. We detected just one frequency, which is the closest to f37 listed by \citet{reed19}. This is surprising, since we would expect the frequency in the \tess\ data to be one of the highest-amplitude frequencies in the \ktwo\ data. It is thus likely that the pulsation properties of this star change significantly over a few years. We show an amplitude spectrum in Fig.\,\ref{fig:USC_ft3} and list the prewhitened frequencies in Table\,\ref{tab:USCfreq}. 

\begin{figure}
\centering
\includegraphics[width=\hsize]{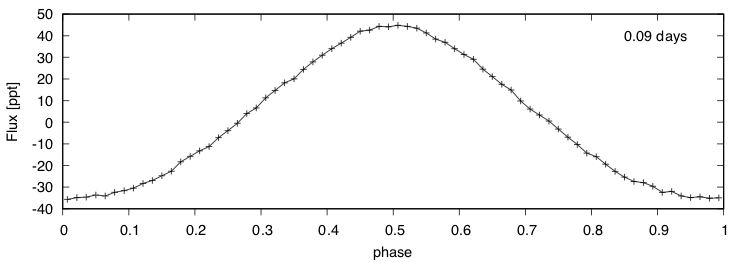}
\caption{Phase-folded light curve of TIC\,409644971 showing a reflection effect. The number in the upper right corner is the orbital period, rounded to two significant digits, used for folding the light curve.}
\label{fig:tic409644971}
\end{figure}

\paragraph{TIC\,355058528}
(Gaia\,DR2\,6692773045444107008, V4640\,Sgr, EC\,20117-4014, CD-40\,13747) is a known sdB pulsator. \citet{odonoghue97} analyzed its spectra and concluded that they are similar to the spectra of other sdB stars. The authors also analyzed photometric time-series data and reported three frequencies, which were later confirmed by \citet{randall06}. The latter author attempted a detailed asteroseismic analysis, arriving at two potential families of optimal models. \tess\ observed the star during Sector\,27. We found only one frequency, which was reported by previous authors as the dominant amplitude. We show an amplitude spectrum in Fig.\,\ref{fig:USC_ft3} and list the prewhitened frequencies in Table\,\ref{tab:USCfreq}.

\begin{figure*}
\centering
\includegraphics[width=\hsize]{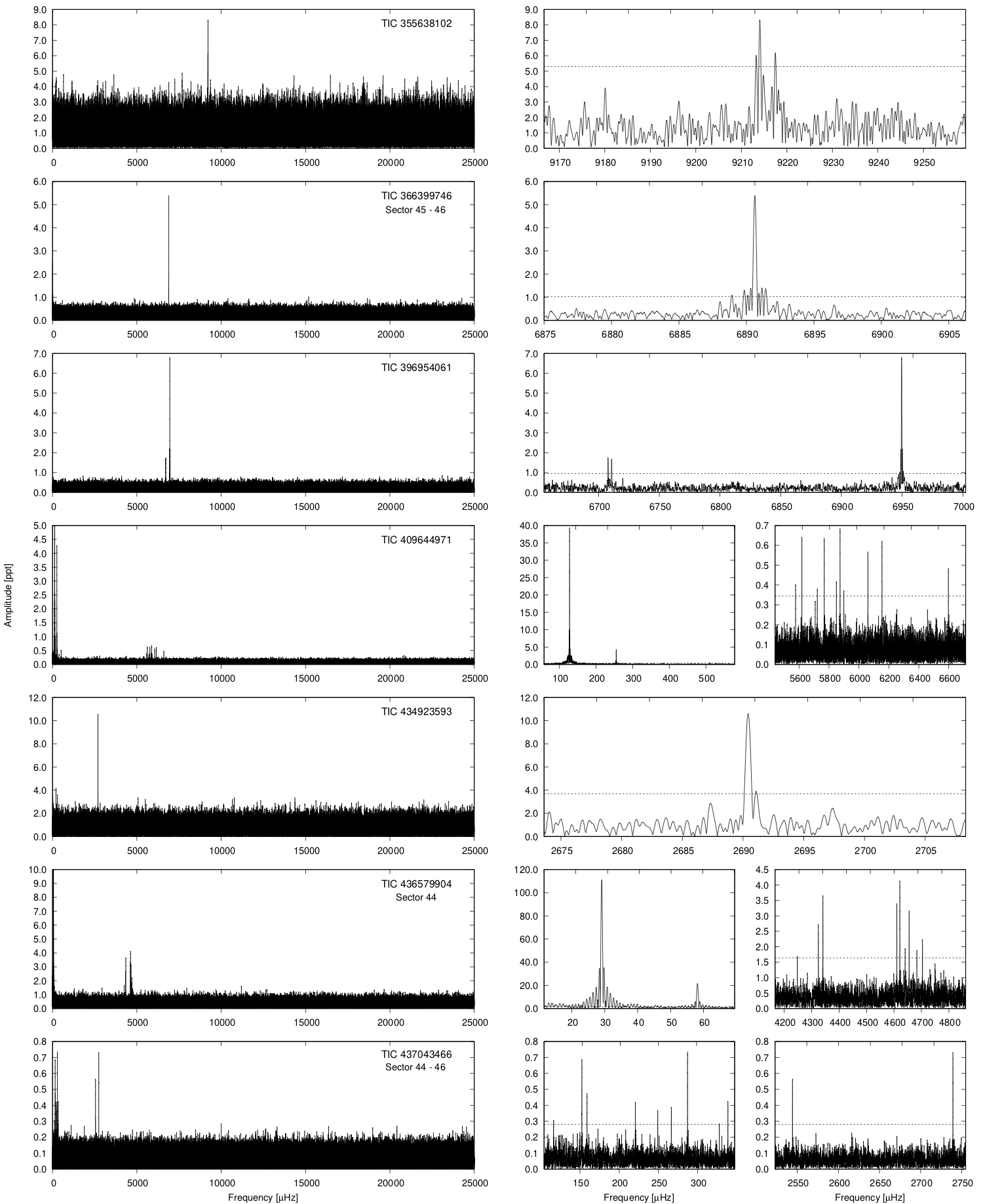}
\caption{Same as in Fig.\,\ref{fig:USC_ft1}, but for another seven targets observed in the USC mode.}
\label{fig:USC_ft4}
\end{figure*}

\paragraph{TIC\,355638102}
(Gaia\,DR2\,6494992795056667648, EC\,23507-5733, LB\,1535) is a new pulsator. \citet{kilkenny16a} classified it as a He-sdO star. Based on a spectrum taken with the 1.9 m telescope at SAAO, we classified it as a He-sdOB star. \tess\ observed the star during Sector\,28. We detected only two close frequencies. We show an amplitude spectrum in Fig.\,\ref{fig:USC_ft4} and list the prewhitened frequency in Table\,\ref{tab:USCfreq}.

\paragraph{TIC\,366399746}
(Gaia\,DR2\,601188910547673728) is a new sdB pulsator. The star has been classified as an sdB+MS binary system by \citet{luo16}. \citet{sahoo20a} reported one low frequency, interpreting it as a binary signature, and consequently the object was proposed and successfully observed by \tess\ during Sectors\,45 and 46. We confirmed the low frequency, but it is of a very low amplitude. We detected one significant high frequency. We show an amplitude spectrum in Fig.\,\ref{fig:USC_ft4} and list the prewhitened high frequency in Table\,\ref{tab:USCfreq}.

\paragraph{TIC\,396954061}
(Gaia\,DR2\,3259060049366022400, 2M\,0415+0154) is a known sdB pulsator. \citet{oreiro09} analyzed its spectra and derived \teff\,=\,34\,000(500)\,K and \loggcms\,=\,5.80(5), which are characteristic of sdB stars. They also analyzed time-series data and reported three frequencies. \tess\ observed the star during Sector\,32. We confirm $f_1$ listed by \citet{oreiro09}, while the other two frequencies we found differ by a few \uHz; these two are close together and could have not been resolved well in the run obtained by \citet{oreiro09}. We show an amplitude spectrum in Fig.\,\ref{fig:USC_ft4} and list the prewhitened frequencies in Table\,\ref{tab:USCfreq}.

\paragraph{TIC\,409644971}
(Gaia\,DR2\,5947131955116293760) is a new sdB pulsator. \citet{nemeth12} analyzed the star spectroscopically and concluded that it is a binary consisting of an sdB star and an F7 main-sequence component. \tess\ observed the star during Sector\,39. As with TIC\,142200764, the light curve shows a characteristic reflection effect with a period of 0.09\,d and we show the phase-folded data in Fig.\,\ref{fig:tic409644971}, while the ephemeris is provided in Table\,\ref{tab:ephemerides}. The binary frequency and its harmonics are listed in Table\,\ref{tab:USCfreq}. The reflection effect must be caused by a companion of lower mass than the sdB star, and so it is typically an M-dwarf. The F7 star visible in the optical spectrum is most likely in a wide orbit, or is possibly unbound from the sdB+dM binary. Follow-up low-resolution spectroscopy would be useful to confirm the orbit of the sdB+dM binary, and high-resolution spectroscopy is required to establish if the F star is orbiting the inner binary. Hierarchical triple systems may play an essential role in forming many sdB stars, including the many apparently single stars that are observed, as explored in the recent paper by \citet{preece22}. According to their analysis, while the majority of the initial triple systems end up in configurations where the outer tertiary escapes after the mass-loss episode that produces the sdB star, either as a merger remnant or as a post-common-envelope binary, many are predicted to remain in a triple configuration. In addition to the binary signal, we detected ten frequencies in the high-frequency region, which we associate with p modes. We show an amplitude spectrum in Fig.\,\ref{fig:USC_ft4}. If these frequencies turn out to be phase stable over time, there is a chance that the orbit of the sdB+dM binary around the F star can be explored through the light-travel-time effect.

\paragraph{TIC\,434923593}
(Gaia\,DR2\,5862700251105587968, CS\,1246) is a known sdB pulsator. \citet{barlow10} confirmed the sdB type and reported on their analysis of photometric data and the discovery of one large-amplitude frequency. \tess\ observed the star during Sector\,38. We detected one high frequency with an amplitude of 10.6\,ppt. The star is known for a large decrease in the amplitude of its only frequency, which is confirmed with our detection, that is to say the pulsation amplitude is about 30\% of the discovery amplitude, and is consistent with the amplitude recently observed from the ground (unpublished). We show an amplitude spectrum in Fig.\,\ref{fig:USC_ft4} and we list the prewhitened frequency in Table\,\ref{tab:USCfreq}.

While checking on the contamination of TIC\,434923593, we found TIC\,434923595 and TIC\,434925198 to be variable stars. We retrieved the light curves from the SC data and calculated the amplitude spectra. There is only one frequency in each star detected. In TIC\,434923595 we found 203.674(32)\,\uHz\ at 2.18(29)\,ppt (S/N\,=\,6.3), while in TIC\,434925198 we found 202.291(39)\,\uHz\ at 1.12(19)\,ppt (S/N\,=\,5.1).

\begin{figure}
\centering
\includegraphics[width=\hsize]{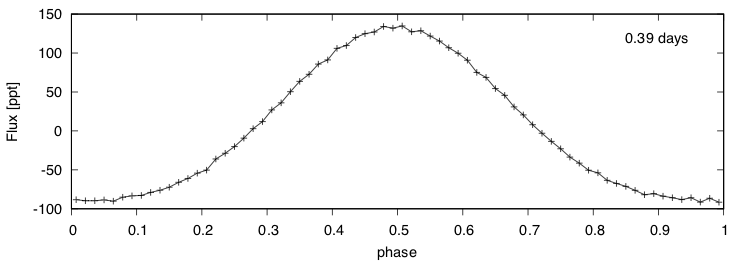}
\caption{Phase-folded light curve of TIC\,436579904 showing a reflection effect. The number in the upper right corner is the orbital period, rounded to two significant digits, used for folding the light curve.}
\label{fig:tic436579904}
\end{figure}

\paragraph{TIC\,436579904}
(Gaia\,DR2\,3308791681845675136, V1405\,Ori, KUV\,04421+1416) is a known sdB pulsator. The \citet{kilkenny88} catalog lists it an sdB star, and \citet{koen99} analyzed time-series data and reported seven frequencies. \citet{reed20} reported on their analyses of \ktwo\ data, which detected a flux variation caused by binarity, confirming the suspicion of \citet{koen99}. The flux variation is caused by the reflection effect and the orbital period is 0.398\,d. \citet{reed20} listed 107 high frequencies attributed to p modes. Some of the frequencies were explained by rotationally split multiplets. Additionally, 19 frequencies were detected in the low-frequency region and attributed to g modes. The authors used an asymptotic period spacing to identify the modal degree. \tess\ observed the star during Sectors\,32, 43, and 44. The phase-folded light curve showing the reflection effect is presented in Fig.\,\ref{fig:tic436579904}, while the ephemeris is given in Table\,\ref{tab:ephemerides}. First, we calculated amplitude spectra from single-sector data. In each spectrum we detected a handful of high frequencies. However, they vary significantly between sectors, and therefore we decided to analyze the sector data separately. We show an amplitude spectrum calculated only from Sector\,44 data in Fig.\,\ref{fig:USC_ft4} and list the prewhitened frequencies in Table\,\ref{tab:USCfreq}.

\paragraph{TIC\,437043466}
(Gaia\,DR2\,611587919724027392, GALEX J085649.3+170115, EPIC\,211779126) is a known sdB pulsator, identified as an sdB star by \citet{nemeth12}. Pulsations were found by \citet{baran17}, who reported the results of their analysis of the \ktwo\ data. They found 154 frequencies in the g-mode region and 29 in the p-mode region, which makes the star a rich hybrid pulsator. Rotationally split modes allowed for an estimation of a
rotation period of $\sim$16\,d to be made. Trapped modes were also reported. \tess\ observed the star during Sectors\,44, 45, and 46. First, we compared the amplitude spectra calculated from single-sector data. We detected two frequencies in the p-mode region in all sectors, which are the highest-amplitude frequencies found by \citet{baran17}. The g-mode region looks different in each sector, but the majority of frequencies repeat. When we combined all sector data together, we detected all but one frequency, f$_{\rm 3}$ in Sector\,45, which had been detected in single-sector data; on the other hand, we detected two frequencies, f$_{\rm 6}$ and f$_{\rm 7}$ in Sectors\,44-46, which are not significant in single-sector data. In Table\,\ref{tab:USCfreq} we show all the frequencies we found, while in Fig.\,\ref{fig:USC_ft4} we show an amplitude spectrum calculated from merged data.

\begin{figure*}
\centering
\includegraphics[width=\hsize]{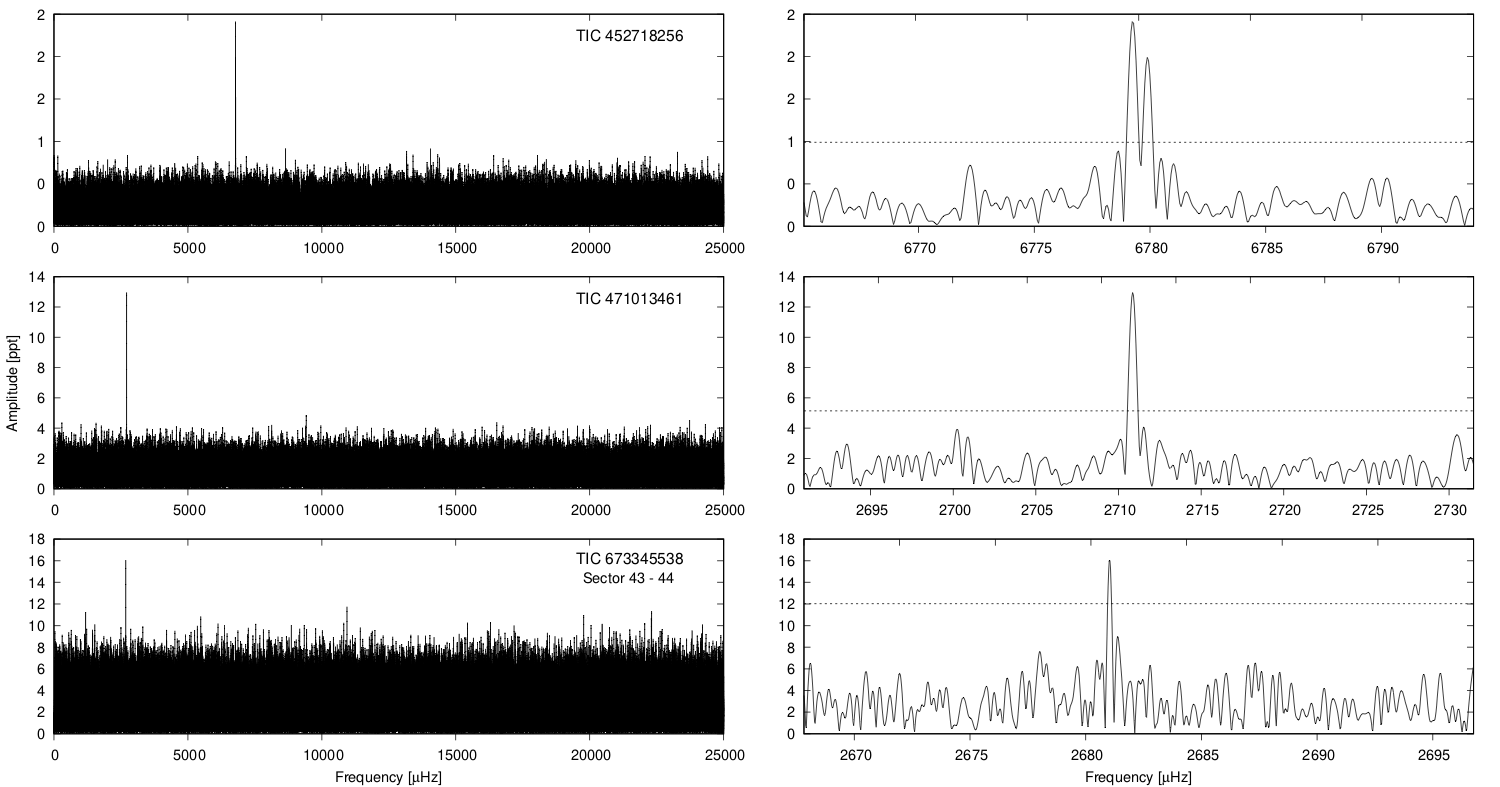}
\caption{Same as in Fig.\,\ref{fig:USC_ft1}, but for another three targets observed in the USC mode.}
\label{fig:USC_ft5}
\end{figure*}

\paragraph{TIC\,452718256}
(Gaia\,DR2\,3533727090595727104, EC\,11275-2504) is a known sdOB pulsator that was classified as an sdOB star by \citet{kilkenny97b}. \citet{kilkenny19} reported the discovery of a signal at frequencies around 6.78\,mHz, and between 7.5 and 7.8\,mHz, depending on data taken sporadically over 2\,years. \tess\ observed the star during Sector\,36, from which we detected two close frequencies around 6779\,\uHz. We show an amplitude spectrum in Fig.\,\ref{fig:USC_ft5} and list the prewhitened frequencies in Table\,\ref{tab:USCfreq}.

\paragraph{TIC\,471013461}
(Gaia\,DR2\,4843383737223292672, EC\,03530-4024) is a new sdB pulsator. \citet{kilkenny16a} classified it as an sdB star. \tess\ observed the star during Sector\,31, from which we detected only one frequency in the p-mode region. We show an amplitude spectrum in Fig.\,\ref{fig:USC_ft5} and list the prewhitened frequency in Table\,\ref{tab:USCfreq}.

\paragraph{TIC\,673345538}
(Gaia\,DR2\,3296244226946647168, V1835\,Ori) is a known sdB pulsator. \citet{ramsay05} found short-period variability in the star, while \citet{ramsay06} classified the as an sdB star and identified the flux variation as p-mode pulsations. The low \teff\ of 29\,200(1\,900)\,K correlates with pulsation period below 3\,000\,\uHz. \citet{baran11b} reobserved the star over the course of 47\,d from two different sites and reported six frequencies in the g-mode region and six frequencies, including one combination, in the p-mode region. That discovery makes the star a hybrid sdB pulsator. \tess\ observed the star during Sectors\,43 and 44. We detected only one frequency, which is the highest-amplitude frequency reported by \citet{baran11b}. We show an amplitude spectrum in Fig.\,\ref{fig:USC_ft5} and list the prewhitened frequency in Table\,\ref{tab:USCfreq}.

\section{Summary}
We present the results of our search for short-period pulsating hot subdwarfs. We used \tess\ space data collected mostly during years\,1 and 3, covering the southern ecliptic hemisphere. We report the stars that show signals at high frequencies, which we associate with p-mode pulsations. In most cases the amplitudes of p modes dominate amplitude spectra. However, there are two stars that have gravity modes of comparable amplitude. Except for those two stars, 41 hot subdwarfs in our sample can be considered p-mode-dominated pulsators. In total, we detected short-period pulsations in 43 hot-subdwarf stars. We found 32 (17 new) sdB stars, eight (three new) sdOB stars, two sdO stars, and one new He-sdOB star. 

We expect that the new hot-subdwarf p-mode pulsators presented in this paper will trigger follow-up monitoring in the future, either from the ground or from space (e.g., PLATO or further TESS observations). Moreover, in cases where the number of frequencies detected with TESS is not higher than what was seen previously from the ground, there is still useful information on the confirmed frequencies: there is no more ambiguity caused by daily aliases, and the precision is higher due to the long time baseline and coverage obtained by TESS. In addition, observing the same frequencies at different epochs can be useful for $\dot{P}$ analyses (e.g., linked to the evolution of the star) or O-C studies (e.g., linked to the presence of a yet undetected companion).

We prewhitened pulsation frequencies above an assumed detection threshold from the \tess\ light curves. The number of frequencies detected in a given star varies from one to 42. Some stars show combination frequencies and/or low frequencies along with harmonics. The latter are a typical signature of binarity. We identified three stars showing large-amplitude binary signatures and we phased the data to show the orbital variation. For TIC\,156618553, which is the well-known prototype of the HW\,Vir-type eclipsing binaries, we derived the mid-times of two \tess-orbit averaged eclipses to high accuracy.

We looked for multiplets that may help in identifying the geometry of the modes and  estimated rotation periods of those pulsators. Out of our 43 targets, multiplets had been previously reported for only six known pulsators: TIC\,47377536, TIC\,60257911, TIC\,62483415, TIC\,335635628, TIC\,436579904, and TIC\,437043466. We can confirm none of those multiplets. In four cases, two new (TIC\,139481265 and TIC\,241771689) and two known (TIC\,69298924 and TIC\,220573709) pulsators, we found candidates for rotationally split frequencies. However, these detections should be confirmed with longer baseline data. The estimated rotation periods range between 1 and 12.9\,d.

As mentioned in the Introduction, pulsations in sdO stars are extremely rare and it is of great importance to expand the numbers of these stars in order to establish the boundaries within which pulsations can be driven in hot-subdwarf stars. We will present in a forthcoming paper (that will include p-mode pulsators detected in the northern ecliptic hemisphere) a detailed statistical analysis on the presence of p-mode pulsators among hot subdwarfs, and their boundaries in a $T_{\rm eff}$-$\log g$ diagram.

While no new Helium-poor sdO stars were found to display pulsations in the \tess\ data, the discovery of two significant pulsation modes in TIC\,355638102 is very surprising. Spectroscopically, this is a iHe-sdOB star and it is roughly comparable to the V366\,Aqr pulsators. A rough assessment of the available spectrum indicates a \teff\ around 40\,kK and a surface gravity in the vicinity of \logg\,=\,6, which is slightly on the hot side of the V366 Aqr stars. All the pulsators of this class have been found to have unusual heavy-metal abundances in their atmospheres, in particular extreme levels of Ge, Sr, Y, and Zr. A high-resolution spectrum of the star would be required in order to establish if this is also the case for TIC\,355638102. However, the V366\,Aqr stars show pulsations in the range of 2000 -- 6000\,s, which must be g modes in subdwarf stars, and the pulsations we detect in TIC\,355638102 are at 108.5\,s, which can only be p modes. A pulsation mechanism invoking the enhancement of carbon was proposed by \citet{saio19} and further explored by \citet{bertolami22}, and has had some success in explaining the presence of g modes at such high temperatures, but has not attempted to explain the absence of p modes in the V366\,Aqr pulsators. Further photometric and spectroscopic studies of TIC\,355638102 are required to make a clear connection here.

Besides the primary goal of this paper, we also found three other new variables, which happened to be located in the target masks of two of our program targets. We detected these stars while doing contamination analysis to avoid false positives. The frequency detected in TIC\,434925198 is only marginally above the detection threshold, while the signals in TIC\,434923595 and TIC\,754255960 are significant. The latter shows three harmonics, indicating a non-sinusoidal shape of the flux variation. This can be interpreted as, for example, shallow eclipses or radial pulsations in classical pulsators. More precise photometric and spectroscopic observations are required to classify the variability types of these three new variables.

\begin{acknowledgements}
We thank Don Kurtz for valuable comments, which have significantly improved the quality of the manuscript. Financial support from the National Science Centre in Poland under project No.\,UMO-2017/26/E/ST9/00703 is acknowledged. SC acknowledges financial support from the Centre National d’\'Etudes Spatiales (CNES, France) and from the Agence Nationale de la Recherche (ANR, France) under grant ANR-17-CE31-0018. PN acknowledges support from the Grant Agency of the Czech Republic (GA\v{C}R 22-34467S). The Astronomical Institute in Ond\v{r}ejov is supported by the project RVO:67985815. This paper includes data collected with the {\it TESS} mission, obtained from the MAST data archive at the Space Telescope Science Institute (STScI). Funding for the {\it TESS} mission is provided by the NASA Explorer Program. STScI is operated by the Association of Universities for Research in Astronomy, Inc., under NASA contract NAS 5–26555. This paper uses observations made at the South African Astronomical Observatory (SAAO). This paper uses observations made at the Nordic Optical Telescope (NOT). This research has made use of the SIMBAD database, operated at CDS, Strasbourg, France. This work has also made use of data from the European Space Agency (ESA) mission {\it Gaia} (\url{https://www.cosmos.esa.int/gaia}), processed by the {\it Gaia} Data Processing and Analysis Consortium (DPAC, \url{https://www.cosmos.esa.int/web/gaia/dpac/consortium}). Funding for the DPAC has been provided by national institutions, in particular the institutions participating in the {\it Gaia} Multilateral Agreement. V.V.G. is a F.R.S.-FNRS Research Associate. This research has used the services of \url{www.Astroserver.org}.
\end{acknowledgements}


\bibliographystyle{aa}
\bibliography{myrefs}

\end{document}

%% file: targets_table.tex
\begin{table}
\centering
\caption{The list of pulsating hot subdwarfs found in the \tess\ data. No USC data were available prior to Sector\,27, hence data in sectors marked in parentheses were not used in our analysis. New pulsators are indicated with \textbf{bold} font.}
\label{tab:targets_all}
\begin{tabular}{|c|r|r|c|}
\hline
&\multicolumn{1}{c|}{TIC} & \multicolumn{1}{c|}{Sector} & \multicolumn{1}{c|}{SpT} \\
\hline
\multirow{11}{*}{\rotatebox[origin=c]{90}{SC targets}}
& \textbf{10011123} & 33 & sdOB \\
& \textbf{19690565} & 34 & sdB \\
&95752908 & 8 & sdB \\
&\textbf{139481265} & 33 & sdB \\
&142200764 & 3 & sdB \\
&\textbf{143699381} & 13 & sdB \\
&\textbf{289149727} & 38 & sdB \\
&\textbf{295046932} & 39 & sdB \\
&\textbf{366656123} & 34,44 & sdB \\
&\textbf{387107334} & 13 & sdB \\
&\textbf{408147637} & 38 & sdB \\
&\textbf{455095580} & 34 & sdB \\
\hline
\multirow{30}{*}{\rotatebox[origin=c]{90}{USC targets}}
& \textbf{6116091} & 37 & sdOB \\
& \textbf{29840077} & (1),28 & sdB \\
& 33318760 & (8),35 & sdB \\
& 47377536 & (9),35,46 & sdB \\
& \textbf{53826859} & 33 & sdB \\
& 60257911 & 44,45,46 & sdB \\
& 62381958 & (1),30 & sdOB \\
& 62483415 & (1),28 & sdOB \\
& 69298924 & 44,45,46 & sdB \\
& \textbf{70549283} & 44,45,46 & sdOB \\
& 98871628 & (10),36 & sdB \\
&139723188 & (1),27,28 & sdOB \\
&156618553 & 46 & sdB \\
&\textbf{169285097} & (2),29 & sdB \\
&220573709 & (1),(2),(3),28,30 & sdO \\
&\textbf{241771689} & 38 & sdB \\
&248949857 & (3),30 & sdO \\
&\textbf{273218137} & (10),37 & sdB \\
&322009509 & (2),29 & sdOB \\
&335635628 & 46 & sdB \\
&355058528 & 27 & sdB \\
&\textbf{355638102} & (1),(2),28 & He-sdOB \\
&\textbf{366399746} & 45,46 & sdB \\
&396954061 & (5),32 & sdB \\
&\textbf{409644971} & (13),39 & sdB \\
&434923593 & 38 & sdB \\
&436579904 & (5),32,43,44 & sdB \\
&437043466 & 44,45,46 & sdB \\
&452718256 & (9),36 & sdOB \\
&\textbf{471013461} & (4),31 & sdB \\
&673345538 & 43,44 & sdB \\
\hline
\end{tabular}
\end{table}

%% file: SC_table.tex
\begin{table}
\centering
\caption{List of frequencies detected in the targets observed only with the SC.}
\label{tab:SCfreq}
\begin{tabular}{cccrc}
\hline\hline
\multirow{2}{*}{ID} & Frequency & Period & \multicolumn{1}{c}{Amplitude} & \multirow{2}{*}{S/N}\\
& [\uHz] & [sec] & \multicolumn{1}{c}{[ppt]} & \\
\hline\hline
\multicolumn{5}{c}{{\bf TIC\,1001123}}\\
f$_{\rm 1}$ &  207.365(29) &   4822.4(7) &  11.9(1.4) &   7.1\\
f$_{\rm 2}$ &  267.617(25) &   3736.68(36) &  13.3(1.4) &   8.0\\
f$_{\rm 3}$ &  293.892(37) &   3402.62(43) &   9.1(1.4) &   5.4\\
f$_{\rm 4}$ & 5794.3287(47) &  172.58255(14) &  72.4(1.4) &  43.2\\
\hline
\multicolumn{5}{c}{{\bf TIC\,19690565}}\\
f$_{\rm 1}$ & 4440.736(38) &    225.1879(19) &   3.3(5) &   5.5\\
f$_{\rm 2}$ & 4464.242(47) &    224.0022(23) &   2.8(5) &   4.5\\
\hline
\multicolumn{5}{c}{{\bf TIC\,95752908}}\\
f$_{\rm 1}$ & 7122.958(24) &    140.39110(48) &   0.49(5) & 8.1\\
f$_{\rm 2}$ & 7369.485(6) &    135.69469(11) &   2.06(5) & 34.4\\
\hline
\multicolumn{5}{c}{{\bf TIC\,139481265}}\\
f$_{\rm 1}$ & 6320.398(29) &    158.2179(7) &   1.12(13) &   7.3\\
f$_{\rm 2}$ & 6321.948(14) &    158.17908(35) &   2.45(14) &  15.8\\
f$_{\rm 3}$ & 6323.546(45) &    158.1391(11) &   0.72(13) &   4.7\\
\hline
\multicolumn{5}{c}{{\bf TIC\,142200764}}\\
f$_{\rm orb}$ &   25.7194(9) &  38881.2(1.4) &  74.81(24) & 266.2\\
2f$_{\rm orb}$ &   51.438743 &  19440.599451 &  15.06(24) &  53.6\\
f$_{\rm 1}$ & 3227.112(36) &    309.8746(34) &   2.06(24) &   7.3\\
f$_{\rm 2}$ & 3463.071(18) &    288.7611(15) &   4.15(24) &  14.8\\
f$_{\rm 3}$ & 3465.906(18) &    288.5248(15) &   4.02(24) &  14.3\\
f$_{\rm 4}$ & 3486.66(7) &    286.807(5) &   1.12(24) &   4.0\\
f$_{\rm 5}$ & 3511.989(33) &    284.7389(27) &   2.21(24) &   7.9\\
f$_{\rm 6}$ & 3516.297(40) &    284.3901(33) &   1.82(24) &   6.5\\
f$_{\rm 7}$ & 3543.983(27) &    282.1684(22) &   2.72(24) &   9.7\\
f$_{\rm 8}$ & 3561.62(6) &    280.771(5) &   1.15(24) &   4.1\\
\hline
\multicolumn{5}{c}{{\bf TIC\,143699381}}\\
f$_{\rm 1}$ &  204.923(24) &   4879.9(6) &   2.38(24) &   8.3\\
f$_{\rm 2}$ &  311.308(16) &   3212.25(17) &   3.53(24) &  12.4\\
f$_{\rm 3}$ &  320.615(30) &   3119.01(29) &   1.94(24) &   6.8\\
f$_{\rm 4}$ & 5461.957(33) &    183.0846(11) &   1.90(24) &   6.7\\
f$_{\rm 5}$ & 5462.990(34) &    183.0500(11) &   1.85(24) &   6.5\\
\hline
\multicolumn{5}{c}{{\bf TIC\,289149727}}\\
f$_{\rm 1}$ &   50.769(20) &  19696.9(7.6) &  29.9(2.5) &  10.3\\
f$_{\rm 2}$ & 5527.456(12) &    180.91506(39) &  49.9(2.5) &  17.2\\
f$_{\rm 3}$ & 5529.050(9) &    180.86288(30) &  65.5(2.5) &  22.5\\
f$_{\rm 4}$ & 5546.945(19) &    180.2794(6) &  31.2(2.5) &  10.7\\
\hline
\multicolumn{5}{c}{{\bf TIC\,295046932}}\\
f$_{\rm 1}$ & 5524.300(38) &    181.0184(12) &   8.7(1) &   5.1\\
f$_{\rm 2}$ & 5551.400(14) &    180.13473(46) &  23.4(1) &  13.7\\
f$_{\rm 3}$ & 6059.790(19) &    165.0222(5) &  17.5(1) &  10.2\\
f$_{\rm 4}$ & 6177.657(12) &    161.87367(31) &  27.9(1) &  16.3\\
f$_{\rm 5}$ & 6229.586(32) &    160.5243(8) &  10.4(1) &   6.0\\
f$_{\rm 6}$ & 6208.480(10) &    161.07001(26) &  34.4(1) &  20.1\\
f$_{\rm 7}$ & 6207.534(12) &    161.09456(30) &  29.0(1) &  16.9\\
f$_{\rm 8}$ & 6210.064(33) &    161.0289(8) &  10.3(1) &   6.0\\
f$_{\rm 9}$ & 6260.176(14) &    159.73992(36) &  23.5(1) &  13.7\\
f$_{\rm 10}$ & 6358.416(39) &    157.2719(10) &   8.5(1) &   5.0\\
f$_{\rm 11}$ & 6452.331(26) &    154.9828(6) &  12.5(1) &   7.3\\
\hline
\multicolumn{5}{c}{{\bf TIC\,366656123 - Sector 34}}\\
f$_{\rm 1}$ &   34.384(6) &  29083.2(4.7) &  15.47(35) &  38.0\\
f$_{\rm 2}$ & 3430.220(33) &    291.5265(28) &   2.62(35) &   6.4\\
\hline
\multicolumn{5}{c}{{\bf TIC\,387107334}}\\
f$_{\rm 1}$ & 5230.956(23) &    191.1696(8) &   3.64(38) &   8.1\\
\hline
\multicolumn{5}{c}{{\bf TIC\,408147637}}\\
f$_{\rm 1}$ &   24.963(23) &   40059(37) &   5.2(5) &   8.6\\
f$_{\rm 2}$ &  323.644(41) &   3089.81(39) &   3.0(5) &   4.9\\
f$_{\rm 3}$ & 5340.857(28) &    187.2359(10) &   4.3(5) &   7.1\\
f$_{\rm 4}$ & 5389.176(33) &    185.5571(11) &   3.7(5) &   6.0\\
f$_{\rm 5}$ & 5537.739(29) &    180.5791(10) &   4.2(5) &   6.8\\
\hline
\multicolumn{5}{c}{{\bf TIC\,455095580}}\\
f$_{\rm 1}$ & 4461.092(36) &    224.1604(18) &   4.1(6) &   5.8\\
\hline\hline
\end{tabular}
\end{table}

%% file: ephemerides.tex
\begin{table}
\centering
\caption{Ephemerides for three stars showing a reflection effect.}
\label{tab:ephemerides}
\begin{tabular}{|r|l|l|}
\hline
\multicolumn{1}{|c}{\multirow{2}{*}{TIC}} & \multicolumn{1}{|c|}{Reference epoch}  & \multicolumn{1}{c|}{Period}\\
& \multicolumn{1}{c|}{[BJD]} & \multicolumn{1}{c|}{[days]}\\
\hline
142200764 & 2458386.7338(10) & 0.450060(39) \\
409644971 & 2458657.06556(34) & 0.090740491(12) \\
436579904 & 2458438.90745(38) & 0.39800589(18) \\
\hline
\end{tabular}
\end{table}